\documentclass[journal, onecolumn]{IEEEtran}

\usepackage{amsthm,amsmath,amssymb,color,fullpage,cite,bm,hyperref,url,cleveref,natbib}
\usepackage{mathrsfs}
\usepackage{graphicx}
\graphicspath{ {./} }
\usepackage[linesnumbered,lined,ruled]{algorithm2e}
\usepackage[noend]{algpseudocode}
\usepackage[margin=1in]{geometry}
\usepackage[singlespacing]{setspace}
\usepackage{dsfont}

\usepackage{xspace}
\usepackage{bbm}
\input{mathlig}
\usepackage{mathtools}

\newcommand{\muspace}{\mspace{1mu}}

\DeclareRobustCommand{\scond}{\mathchoice{\muspace\vert\muspace}{\vert}{\vert}{\vert}}
\mathlig{|}{\scond}
\newcommand{\cond}{\mathchoice{\,\vert\,}{\mspace{2mu}\vert\mspace{2mu}}{\vert}{\vert}}

\DeclareRobustCommand{\discint}{\mathchoice{\mspace{-1.5mu}:\mspace{-1.5mu}}{\mspace{-1.5mu}:\mspace{-1.5mu}}{:}{:}}
\mathlig{::}{\discint}

\def\eps{\epsilon}

\DeclareMathOperator\E{\mathsf{E}}
\let\P\relax
\DeclareMathOperator\P{\mathsf{P}}

\def\textiid{i.i.d.\@\xspace}
\newcommand\iid{\ifmmode\text{ i.i.d. } \else \textiid \fi}

\def\mathllap{\mathpalette\mathllapinternal}
\def\mathllapinternal#1#2{%
  \llap{$\mathsurround=0pt#1{#2}$}}

\def\clap#1{\hbox to 0pt{\hss#1\hss}}
\def\mathclap{\mathpalette\mathclapinternal}
\def\mathclapinternal#1#2{%
  \clap{$\mathsurround=0pt#1{#2}$}}

\let\oldstackrel\stackrel
\renewcommand{\stackrel}[2]{\oldstackrel{\mathclap{#1}}{#2}}

\DeclarePairedDelimiterX{\divx}[2]{(}{)}{%
  #1\delimsize\|#2%
}

\renewcommand{\hbar}{h\mathllap{\overline{\vphantom{h}\hphantom{\rule{4.6pt}{0pt}}}\mspace{0.77mu}}}

\catcode`~=11 
\newcommand{\urltilde}{\kern -.06em\lower -.06em\hbox{~}\kern .02em}
\catcode`~=13 

\usepackage{amsfonts}
\usepackage{dsfont}

\DeclarePairedDelimiter{\abs}{\lvert}{\rvert}

\newtheorem{theorem}{Theorem}
\newtheorem{corollary}{Corollary}
\newtheorem{proposition}{Proposition}
\newtheorem{lemma}{Lemma}

\theoremstyle{definition}

\newtheorem{remark}{Remark}

\newcommand{\sfth}{\mathsf{TH}}

\def\DKL{{D_\mathrm{KL}(p\|1-p)}}
\def\div{{D_\mathrm{KL}}}
\def\Binom{{\mathrm{Binom}}}
\def\eps{{\epsilon}}
\def\cs{{\mathcal S}}
\def\ct{{\mathcal T}}
\def\ce{{\mathcal E}}
\def\ca{{\mathcal A}}
\def\cb{{\mathcal B}}
\def\bx{{\mathbf x}}

\def\lb{{\alpha(p-\eps)}}

\newcommand{\orn}{\ensuremath{\mathsf{OR}} }
\newcommand{\andn}{\ensuremath{\mathsf{AND}} }
\newcommand{\maxn}{\ensuremath{\mathsf{MAX}} }
\newcommand{\constructheap}{\textsc{ConstructMaxHeap} }
\title{Noisy Computing of the Threshold Function}

\author{Ziao Wang, Nadim Ghaddar, Banghua Zhu, and Lele Wang
\thanks{Ziao Wang is with the Department of Electrical and Computer Engineering, University of British Columbia, Vancouver, BC V6T 1Z4, Canada (email: ziaow@ece.ubc.ca).}
\thanks{Nadim Ghaddar is with the Department of Electrical and Computer Engineering, University of Toronto, Toronto, ON M5S 3G8, Canada, (email: nadim.ghaddar@utoronto.ca).}
\thanks{Banghua Zhu is with the Department of Electrical Engineering and Computer Sciences, University of California Berkeley, Berkeley, CA 94720, USA, (email: banghua@cs.berkeley.edu).}
\thanks{Lele Wang is with the Department of Electrical and Computer Engineering, University of British Columbia, Vancouver, BC V6T 1Z4, Canada (email: lelewang@ece.ubc.ca).}}

\date{}

\begin{document}
\allowdisplaybreaks

\maketitle
\begin{abstract}
  Let $\sfth_k$ denote the $k$-out-of-$n$ threshold function: given $n$ input Boolean variables, the output is $1$ if and only if at least $k$ of the inputs are $1$. We consider the problem of computing the $\sfth_k$ function using noisy readings of the Boolean variables, where each reading is incorrect with some fixed and known probability $p \in (0,1/2)$. As our main result, we show that it is sufficient to use
        $(1+o(1)) \frac{n\log \frac{m}{\delta}}{D_{\mathsf{KL}}(p \| 1-p)}$
    queries in expectation to compute the $\sfth_k$ function with a vanishing error probability $\delta = o(1)$, where $m\triangleq \min\{k,n-k+1\}$ 
    and $D_{\mathsf{KL}}(p \| 1-p)$ denotes the Kullback-Leibler divergence between $\mathsf{Bern}(p)$ and $\mathsf{Bern}(1-p)$ distributions. Conversely, we show that any algorithm achieving an error probability of $\delta = o(1)$ necessitates at least $(1-o(1))\frac{(n-m)\log\frac{m}{\delta}}{D_{\mathsf{KL}}(p \| 1-p)}$ queries in expectation.
   The upper and lower bounds are tight when $m=o(n)$, and are within a multiplicative factor of $\frac{n}{n-m}$ when $m=\Theta(n)$. In particular, when $k=n/2$, the $\sfth_k$ function corresponds to the $\mathsf{MAJORITY}$ function, in which case the upper and lower bounds are tight up to a multiplicative factor of two.
    Compared to previous work, our result tightens the dependence on $p$ in both the upper and lower bounds.
\end{abstract}

\section{Introduction}\label{sec:introduction}
Coping with noise in computing is an important problem to consider in large systems. With applications in fault tolerance~\citep{Hastad1987,Pease1980,pippenger1991lower}, active ranking~\citep{Shah2018,Agarwal2017,Falahatgar2017,Heckel2019,wang2024noisy,gu2023optimal}, noisy searching~\citep{Berlekamp1964,Horstein1963,Burnashev1974,pelc1989searching, Karp2007}, among many others, the goal is to devise algorithms that are robust enough to detect and correct the errors that can happen during the computation. 
More concretely, the problem can be defined as follows: suppose an agent is interested in computing a function $f$ of $n$ variables with an error probability at most $\delta$, as quickly as possible. To this end, the agent can ask binary questions (referred to hereafter as \emph{queries}) about the variables at hand. The binary response to the queries is observed by the agent through a binary symmetric channel (BSC) with crossover probability $p$. The agent can adaptively design subsequent queries based on the responses to previous queries. The goal is to characterize the relation between $n$, $\delta$, $p$, and the query complexity, which is defined as the minimal number of queries that are needed by the agent to meet the intended goal of computing the function $f$.

This paper considers the computation of the threshold-$k$ function. For $n$ Boolean variables $\mathbf{x}=(x_1,\ldots,x_n) \in \{0,1\}^n$, the threshold-$k$ function $\sfth_k(\cdot)$ computes whether the number of 1's in $\mathbf{x}$ is at least $k$ or not, i.e., 
\begin{equation*}
    \sfth_k(\mathbf{x})\triangleq\begin{cases}
        1 & \text{ if } \sum_{i=1}^n x_i\ge k;\\
        0 & \text{ if } \sum_{i=1}^n x_i<k.
    \end{cases}
\end{equation*}
The noisy queries correspond to noisy readings of the bits, where at each time step, the agent queries one of the bits, and with probability $p$, the wrong value of the bit is returned. It is assumed that the constant $p\in(0,1/2)$ is known to the agent. Our goal is to characterize the optimal query complexity for computing the $\sfth_k$ function with error probability at most $\delta$. Throughout the paper, we assume parameters $k$ and $\delta$ are functions that scale with $n$, while $p$ is an absolute constant independent of all other parameters. We also assume that $\delta$ is always strictly positive.

This model for noisy computation of the $\sfth_k$ function has been considered in~\citet{feige1994computing}, where upper and lower bounds on the number of queries that are needed to compute the two functions are derived in terms of the number of variables $n$, the threshold $k$, the noise probability $p$ and the desired error probability $\delta$. The upper and lower bounds in~\citet{feige1994computing} are within a constant factor, hence providing the optimal order for the minimum number of queries; however, the exact tight characterization of the optimal number of queries is still open. 

In this paper, we tighten this gap and provide new upper and lower bounds for the computation of the $\sfth_k$ function, which simultaneously improves the existing upper and lower bounds.


The main result of this paper can be stated as follows: for any $1\le k\le n$, there exists an algorithm that computes the $\sfth_k$ function with an error probability at most $\delta = o(1)$, and the algorithm uses at most 
    $$(1+o(1)) \frac{n\log \frac{m}{\delta}}{D_{\mathsf{KL}}(p \| 1-p)}$$
queries in expectation. Here we define $m\triangleq\min\{k,n-k+1\}$ and denote the Kullback-Leibler divergence between $\mathsf{Bern}(p)$ and $\mathsf{Bern}(1-p)$ distributions by  $D_{\mathsf{KL}}(p \| 1-p)$. Conversely, we prove that to achieve an error probability of $\delta=o(1)$, any algorithm must make at least $$(1-o(1))\frac{(n-m)\log\frac{m}{\delta}}{D_{\mathsf{KL}}(p \| 1-p)}$$ queries in expectation. 
When $m=o(n)$, the ratio between these upper and lower bounds is $1+o(1)$, and hence we provide an asymptotically tight characterization for the optimal number of queries. For general $m$, these bounds are tight within a multiplicative factor of $2$.

\begin{figure}[t]
	\centering
	\vspace{-1em}
	\hspace*{-1.25em}
	\includegraphics[scale=0.6]{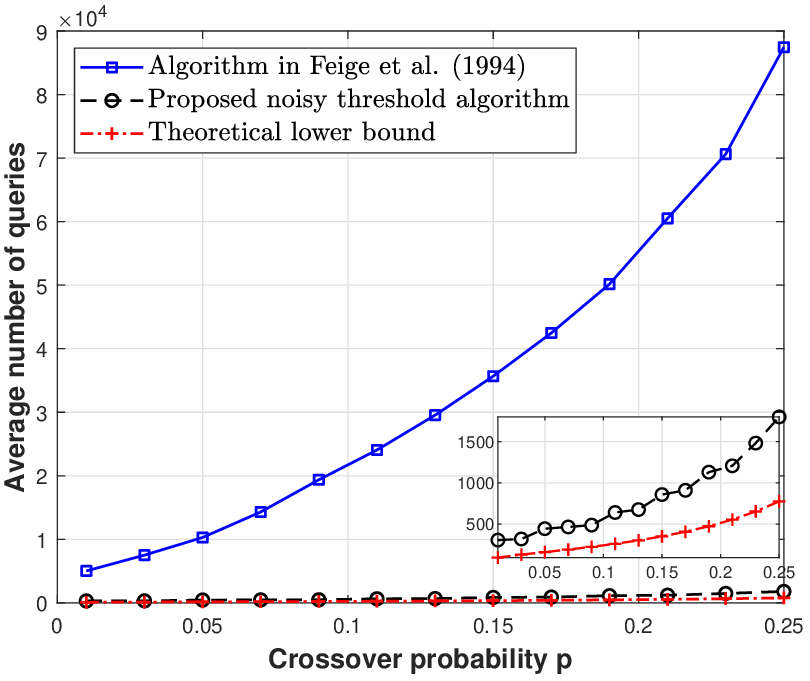}
	\caption{Average number of queries used by the proposed noisy threshold algorithm for computing the $\mathsf{MAJORITY}$ function (i.e., $k=n/2$), in comparison to the algorithm proposed in~\citet{feige1994computing}, for $n=100$ and $\delta=10^{-2}$. The theoretical lower bound corresponds to the plot of $\frac{(n-m)\log(m/\delta)}{\DKL}$, where $m = \min\{k,n-k+1\}$.} 
	\label{fig:comparison}
\end{figure} 

To provide a more quantitative comparison with the bounds presented in~\citet{feige1994computing}, let us consider the example where we set $k=n^{1/3}$, $\delta=n^{-1/4}$, and $p=\frac{1}{3}$.
According to our results, the optimal number of queries to achieve error probability $\delta$ is approximately $2.5247\cdot n\log n$. In contrast, the constants in front of $n\log n$ in the lower and upper bounds given in \citet{feige1994computing} are roughly $0.0506$ and $433.7518$\footnote{The constant in the upper bound of~\citet{feige1994computing} is not explicitly given. We refined the analysis and computed the constant in this example using the algorithm proposed in~\citet{feige1994computing}.}, respectively. As an additional example, Figure~\ref{fig:comparison} shows the simulation of our proposed algorithm in computing the $\mathsf{MAJORITY}$ function (that is, the threshold function with $k=n/2$) compared to the algorithm proposed in~\citet{feige1994computing}, when $n=100$ and $\delta=10^{-2}$. Our algorithm clearly uses fewer queries on average to achieve the desired error probability. Moreover, our algorithm shows a scaling similar to the theoretical lower bound derived in this paper. The detailed setup of the simulation is given in Appendix~\ref{sec:implementation}.

\subsection{Related Work}
\begin{sloppypar}
The noisy computation of the $\sfth_k$ function has primarily been investigated with a focus on the special case of $k=1$, i.e., the \orn function. Early investigations into the noisy computation of the \orn function can be traced back to studies in circuit theory with noisy gates \citep{dobrushin1977lower, dobrushin1977upper, von1956probabilistic, pippenger1991lower, gacs1994lower} and noisy decision trees \citep{evans1998average, reischuk1991reliable}.
For the more general case of arbitrary $k$, the noisy computation of the $\sfth_k$ function was initially explored in the seminal work~\citep{feige1994computing} on noisy computing. Subsequent investigations extended this study to the context of noisy broadcast networks \citep{kushilevitz2005computation}. In the noise model we consider, the best-known upper and lower bounds for the required number of queries to achieve a desired error probability $\delta$ are attributed to~\citet{feige1994computing}. Their work characterizes the optimal order as $\Theta(n\log(k/\delta))$ for the number of queries required. In this work, we improve the known bounds and provide an asymptotically tight characterization for the required number of queries. 
\end{sloppypar}

Recently, there have been several notable efforts to precisely characterize the optimal number of queries with tight constant dependencies for various noisy computing tasks. \citet{wang2024noisy} consider the problem of noisy sorting with pairwise comparisons, and provide upper and lower bounds for the required number of queries with a multiplicative gap of factor $2$. This gap is later closed by~\citet{gu2023optimal}. In a recent paper~\citep{zhu2023ormax}, tight bounds are established for two tasks including computing the \orn function of a binary sequence and \maxn function of a real sequence using pairwise comparisons. 



\subsection{Notation}
In this paper, we adhere to the following notation conventions. Sets and events are represented using calligraphic letters, random variables using uppercase letters, and specific realizations of random variables using lowercase letters. For example, $\mathcal{X}$ and $\mathcal{Y}$ are two sets, $X$ and $Y$ are random variables, and $x$ and $y$ are their respective realizations. Vectors will be denoted using boldface letters, while scalars will be denoted using regular (non-bold) letters. The probability of an event $\mathcal{A}$ will be expressed as $\P(\mathcal{A})$, and the expectation of a real-valued random variable $X$ will be denoted as $\E[X]$. For a finite set $\mathcal{X}$, its cardinality will be denoted by $|\mathcal{X}|$. Unless otherwise specified, we use $\log$ to indicate the natural logarithm. We use $[n]$ to denote the set of integers $\{1,\ldots,n\}$. We follow the standard Landau notation: $f(n) = O(g(n))$ if $\lim_{n \to \infty} \frac{|f(n)|}{g(n)} < \infty$; $f(n) = \Omega(g(n))$ if $\lim_{n \to \infty}\frac{f(n)}{g(n)} >0$; $f(n) = \Theta(g(n))$ if $f(n) = O(g(n))$ and $f(n) = \Omega(g(n))$; $f(n) = o(g(n))$ if $\lim_{n \to \infty}\frac{f(n)}{g(n)} = 0$; $f(n) = \omega(g(n))$ if $\lim_{n \to \infty}\frac{|f(n)|}{|g(n)|} = \infty$; and $f(n) \sim g(n)$ if $\lim_{n \to \infty}\frac{f(n)}{g(n)} = 1$. Unless otherwise specified, limits in all instances of Landau notation are taken with respect to $n$.

\section{Main Results}
The main result of this paper comprises two theorems, through which we provide an asymptotically tight characterization of the minimum expected number of queries required to compute the $\sfth_k$ function with worst-case error probability $\delta$.
\begin{theorem}[Converse]\label{thm:thk-conv}
Suppose $k\le n/2$ and $\delta=o(1)$. Consider any variable-length algorithm for computing $\sfth_k(\mathbf{x})$ that makes $M$ noisy queries. If $M$ satisfies 
    $\E[M|\mathbf{x}]\le (1-o(1))\frac{(n-k)\log\frac{k}{\delta}}{\DKL}$
for any input instance $\mathbf{x}$, then the worst-case error probability of the algorithm is at least $\delta$. 
\end{theorem}
The main technical contribution of this paper is in the proof of Theorem~\ref{thm:thk-conv}. 
In this proof, we consider an enhanced yet more statistically tractable version of any adaptive algorithm for computing the $\sfth_k$ function, and leverage the probabilistic methods to lower bound its error probability.
We illustrate the main idea of the proof in Section~\ref{sec:proof-lower}, while deferring the detailed proof to Appendix~\ref{appd:proof-conv}.

\begin{theorem}[Achievability]\label{thm:threshold_achievability}
    Suppose $k\le n/2$ and $\delta=o(1)$. There exists a variable-length algorithm that computes the $\sfth_k$ function with worst-case error probability at most $\delta$. Moreover, the number of queries $M$ made by the algorithm satisfies
        $\E[M|\bx]\le(1+o(1))\frac{n\log \frac{k}{\delta}}{\DKL}$
    for any input instance $\bx$. 
\end{theorem}
We describe the proposed algorithm, and sketch the proof of Theorem~\ref{thm:threshold_achievability} in Section~\ref{sec:proof-upper}. The detailed proof is presented in Appendix~\ref{appd:proof-ach}.

The upper and lower bounds established in Theorems~\ref{thm:thk-conv} and~\ref{thm:threshold_achievability} are restricted to the case of $k\le n/2$. The following corollary extends the results to any $k\in[n]$.
\begin{corollary}\label{cor:general-k}
    Let $m=\min\{k,n-k+1\}$. For any $k\in [n]$ and $\delta=o(1)$, it is sufficient to use $(1+o(1))\frac{n\log\frac{m}{\delta}}{D_\mathsf{KL}(p\|1-p)}$ queries in expectation to compute the $\sfth_k$ function with worst-case error probability $\delta=o(1)$, while at least $(1-o(1))\frac{(n-m)\log\frac{m}{\delta}}{D_\mathsf{KL}(p\|1-p)}$ queries in expectation are necessary.
\end{corollary}
Corollary~\ref{cor:general-k} follows readily by the fact that $\sfth_k(\bx)=1-\sfth_{n-k+1}(1-\bx)$ for any binary sequence $\bx$, which implies that computing the $\sfth_k$ function is equivalent as computing the $\sfth_{n-k+1}$ function.
\begin{remark}[Gap between the upper and lower bounds]
    The upper and lower bounds presented in Corollary~\ref{cor:general-k} are within a multiplicative factor of $2+o(1)$. To see this, notice that the ratio between the upper and lower bounds in Corollary~\ref{cor:general-k} is $(1+o(1))\frac{n}{n-m}$. Because $m=\min\{k,n-k+1\}\le (n+1)/2$, this ratio has maximum value $2+o(1)$ when $k=(n+1)/2$. The problem of closing this gap remains open.
\end{remark}

\begin{corollary}[Tight bounds for \orn and \andn functions]
\label{cor:OR-AND}
    It is both sufficient and necessary to use 
    $
    (1 \pm o(1)) \frac{n\log \frac{1}{\delta}}{D_{\mathsf{KL}}(p \| 1-p)}
    $
    noisy queries in expectation to compute the \orn and $\mathsf{AND}$ functions of $n$ Boolean variables with worst-case error probability $\delta$.
\end{corollary}
Corollary~\ref{cor:OR-AND} follows by the fact that \orn and \andn functions can be viewed as special cases of the $\sfth_k$ function with $k=1$ and $k=n$ respectively. It recovers the tight bounds for the \orn and $\mathsf{AND}$ functions found by~\cite{zhu2023ormax}.
\begin{corollary}[Fixed-length algorithms]\label{cor:fixed-length}
    Suppose $\delta=o(1)$ and $\delta\ge n^{-1+\eps}$ for some arbitrarily small positive constant $\eps$. Then for any $k\le n/2$, there exists a fixed-length algorithm that computes the $\sfth_k$ function with error probability at most $\delta$. Moreover, the algorithm uses at most $(1+o(1))\frac{n\log\frac{k}{\delta}}{D_\mathsf{KL}(p\|1-p)}$ queries.
\end{corollary}
We prove Corollary~\ref{cor:fixed-length} in Appendix~\ref{appd:proof-ach} by constructing a fixed-length version of the proposed algorithm.

\section{Main Idea in the Proof of the Lower Bound}
\label{sec:proof-lower}
In this section, we present the main idea in the proof of Theorem~\ref{thm:thk-conv}. We illustrate the proof idea for two distinct regimes: $\log k \le \frac{\log(1/\delta)}{\log\log(1/\delta)}$ and $\log k > \frac{\log(1/\delta)}{\log\log(1/\delta)}$, as the techniques applied in these regimes differ significantly.
\subsection{Regime \texorpdfstring{$\log k \le \frac{\log(1/\delta)}{\log\log(1/\delta)}$}{Lg}}
In this regime, our lower bound relies on Le Cam's two-point method.
In this method, we view the noisy computing problem as a binary testing problem. We design an input instance $\bx^{(0)}$ with $\sfth_k(\bx)=0$, and $n-k$ input instances $\bx^{(1)},\ldots,\bx^{(n-k)}$ with $\sfth_k(\bx)=1$ for each $i\in [n-k]$. Using Le Cam's two-point lemma~\citep{yu1997assouad}, we show that when the query complexity of the algorithm is below a certain threshold, the algorithm cannot distinguish $\bx^{(0)}$ and $\bx^{(i)}$ for some $i\in[n-k]$, leading to an estimation error of the $\sfth_k$ function. Specifically, this method yields a lower bound of $(1-o(1))\frac{(n-k)\log\frac1\delta}{\DKL}$ on the required number of queries, and further implies the desired bound $(1-o(1))\frac{(n-k)\log\frac{k}{\delta}}{\DKL}$ by the fact that $\log (k/\delta)=(1+o(1))\log\frac1\delta$ in this regime. We defer the detailed proof in this regime to Appendix~\ref{appd:proof-conv}.

Notably, in the opposite regime $\log k>\frac{\log(1/\delta)}{\log\log(1/\delta)}$, the equality $\log (k/\delta)=(1+o(1))\log\frac1\delta$ does not necessarily hold true, and hence the bound provided by Le Cam's method does not yield the desired lower bound. For example, when $\delta=\frac{1}{\log n}$ and $k=n/2$, the lower bound provided by Le Cam's method is given by $(1-o(1))\frac{\frac{n}{2}\log\log n}{\DKL}$, which does not even match the asymptotic order of the desired lower bound $(1-o(1))\frac{\frac{n}{2}\log \frac{n\log n}{2}}{\DKL}$.

\subsection{Regime \texorpdfstring{$\log k > \frac{\log(1/\delta)}{\log\log(1/\delta)}$}{Lg}}
In this regime with a larger $k$ value, the bound derived from Le Cam's method falls short in capturing the dependency between the optimal number of queries and the parameter $k$. Therefore, we adopt an alternative approach, where we directly analyze the query-response statistics of an \emph{arbitrary} algorithm, and bound its error probability. The major difficulty in establishing a converse result for an arbitrary algorithm is that the algorithm may have an arbitrarily correlated query strategy. To tackle this challenge, we introduce a two-phase enhanced version of any given variable-length algorithm. The first phase of the enhanced algorithm is called the \emph{non-adaptive phase}. It queries each bit for a fixed amount of time. The second phase is called the \emph{adaptive phase}. It adaptively chooses certain bits to \emph{noiselessly} reveal their values. This two-phase design of the enhanced algorithm makes the response statistics more tractable and analyzable. We show that the worst-case probability of error of the enhanced algorithm is less than or equal to that of the original algorithm. Therefore, it suffices to show that the enhanced algorithm has error probability at least $\delta$. 
We sketch the proof in the rest of this section. 

To prove the desired bound $(1-o(1))\frac{(n-k)\log\frac{k}{\delta}}{\DKL}$, we define a vanishing quantity $\epsilon=o(1)$, and prove that any algorithm with at most $\frac{(1-\eps)(n-k-\eps n)\log\frac{k}{\delta}}{\div(p-\epsilon\|1-p)}$ queries in expectation has error probability at least $\delta$.

\emph{\textbf{Enhanced algorithm.}}
Consider an arbitrary algorithm $\mathcal{A}$ that makes at most $\frac{(1-\epsilon)(n-k-\epsilon n)\log \frac{k}{\delta}}{\div(p-\epsilon\|1-p)}$ queries in expectation. We introduce an enhanced version of $\mathcal{A}$, denoted $\mathcal{A}'$, and show that the error probability of $\mathcal{A}'$ is always less than or equal to the error probability of $\mathcal{A}$. We then lower bound the error probability of $\mathcal{A}'$ to complete the proof. 

The enhanced algorithm $\mathcal{A}'$ includes a two-step procedure:
\begin{itemize}
\setlength\itemsep{0.2em}
    \item Non-adaptive phase: query each of the $n$ bits $\alpha\triangleq \frac{(1-\epsilon)\log \frac{k}{\delta}}{\div(p-\epsilon\|1-p)}$ times.
    
    \item Adaptive phase: Adaptively choose $N'$ bits and \emph{noiselessly} reveal their value, where the number of revealed bits $N'$ satisfies $\E[N'|\mathbf{x}]=n-k-\epsilon n$ for any $\mathbf{x}$.
\end{itemize}


The purpose of the enhanced algorithm is to transform an intractable adaptive algorithm into a tractable one with analyzable statistics. The idea of algorithm enhancement is first introduced by~\citet{feige1994computing}, where the authors define an enhancement for any \emph{fixed-length} algorithm. In the above definition of $\ca'$, we extend this notion of algorithm enhancement to accommodate any \emph{variable-length} algorithm, and analyze the error probability for this enhanced variable-length algorithm. Consequently, the lower bound established in our work applies to the expected query complexity of any \emph{variable-length} algorithm, while the lower bound by~\citet{feige1994computing} applies only to the deterministic query complexity of any \emph{fixed-length} algorithm.

It is an intuitive observation that Algorithm $\mathcal{A}'$ works strictly better than $\mathcal{A}$. To see this, we notice that the expected number of bits queried more than $\alpha= \frac{(1-\epsilon)\log \frac{k}{\delta}}{\div(p-\epsilon\|1-p)}$ times in Algorithm $\mathcal{A}$ is at most $n-k-\epsilon n$, because $\mathcal{A}$ makes at most $\frac{(1-\epsilon)(n-k-\epsilon n)\log \frac{k}{\delta}}{\div(p-\epsilon\|1-p)}$ queries in expectation. In contrast, Algorithm $\mathcal{A}'$ makes $\alpha$ queries to each bit, and adaptively choose $n-k-\epsilon n$ bits in expectation to acquire the values noiselessly. As a result, Algorithm $\mathcal{A}'$ obtains more information on every bit than Algorithm $\mathcal{A}$, and hence has a lower error probability. We formally state this result in following lemma, and defer its proof to Appendix~\ref{appd:proof-conv}.
\begin{lemma}
\label{lem:enhancement}
    The worst-case error probability of $\mathcal{A}'$ is at most the worst-case error probability of $\mathcal{A}$.
\end{lemma}

Now, it suffices to show that the worst-case error probability of $\mathcal{A}'$ is at least $\delta$.
To bound the error probability of Algorithm $\mathcal{A}'$, we analyze the query responses from the non-adaptive phase using a balls-and-bins model. We view each of the $n$ bits as a ball with an underlying weight. A bit of value one corresponds to a heavy ball, and a bit of value zero corresponds to a light ball. In the non-adaptive phase, each bit $x_i$ is queried $\alpha$ times. Let $I_i$ denote the number of ones in these $\alpha$ responses. If $I_i = j$, we put ball $i$ to bin $j\in \{0,1,\ldots,\alpha\}$. This creates $\alpha + 1$ bins $\cs_j \triangleq \{i\in[n]:I_i = j\}$ for $j\in \{0,1,\ldots,\alpha\}$. 

The statistics of the responses we get from the non-adaptive phase can be characterized by a sequence of $\alpha+1$ sets $\cs\triangleq(\cs_0,\ldots,\cs_\alpha)$.
For $j\in \{0,\ldots,\alpha\}$, we define $\mathcal{H}_j\triangleq\{i\in[n]:x_i=1, I_i=j\}$ and $\mathcal{L}_j\triangleq\{i\in[n]:x_i=0, I_i=j\}$, i.e., $\mathcal{H}_j$ (resp. $\mathcal{L}_j$) represents the set of heavy (resp. light) balls in the $j$th bin. Let $H_j\triangleq|\mathcal{H}_j|$ and $L_j\triangleq|\mathcal{L}_j|$.  Let ${\bf H}_0^\alpha\triangleq (H_0,\ldots,H_\alpha)$ and ${\bf L}_0^\alpha\triangleq (L_0,\ldots,L_\alpha)$.

In the adaptive phase of Algorithm $\mathcal{A}'$, we noiselessly reveal the weights of $N'$ balls. Using $\cs$ and the weights of the balls revealed in the adaptive phase, Algorithm $\mathcal{A}'$ needs to estimate whether there are at least $k$ heavy balls out of $n$ balls.

\emph{\textbf{Restricting input sequence weights.}}
To facilitate the analysis of Algorithm $\mathcal{A}'$, we restrict the input instance space. We assume that the input sequence $\mathbf{x}$ satisfies either $w(\mathbf{x})=k-1$ or $w(\mathbf{x})=k$, i.e., there are either $k$ or $k-1$ heavy balls. The error for the restricted inputs is a lower bound on the original worst-case error
\[
\max_{\mathbf{x}\in\{0,1\}^n}\P(\widehat{\sfth}_k \neq \sfth_k(\mathbf{x})|\mathbf{x}) \ge \max_{\substack{\mathbf{x}\in\{0,1\}^n:\\w(\mathbf{x}) = k \text{ or}\\ w(\mathbf{x}) = k-1}}\P(\widehat{\sfth}_k \neq \sfth_k(\mathbf{x})|\mathbf{x}),
\]
where $\widehat{\sfth}_k$ denote the output estimate of Algorithm $\ca'$.
Later we will show the error on the restricted input space is at least $\delta$.

\emph{\textbf{Genie-aided information.}}
We assume that a genie provides Algorithm $\mathcal{A}'$ with additional information after the non-adaptive phase to assist the estimation.
Specifically, the genie reveals two sequences of sets $\bar{\mathcal{H}}_{0}^ {\alpha}=(\bar{\mathcal{H}}_{0},\ldots, \bar{\mathcal{H}}_{\alpha})$ and $\bar{\mathcal{L}}_{0}^ {\alpha}=(\bar{\mathcal{L}}_{0},\ldots, \bar{\mathcal{L}}_{\alpha})$, which are defined under two hypotheses $w(\mathbf{x})=k-1$ and $w(\mathbf{x})=k$ respectively as follows. 
\begin{itemize}
\setlength\itemsep{0.2em}
    \item When $w(\mathbf{x})=k-1$, there are in total $k-1$ heavy balls in the bins, and the genie reveals the indices of all of them, i.e., 
        $(\bar{\mathcal{H}}_{0},\ldots, \bar{\mathcal{H}}_{\alpha})=(\mathcal{H}_{0},\ldots, \mathcal{H}_{\alpha})\;\;\text{and}\;\;(\bar{\mathcal{L}}_{0},\ldots, \bar{\mathcal{L}}_{\alpha})=(\mathcal{L}_{0},\ldots, \mathcal{L}_{\alpha}).$
\item When $w(\mathbf{x})=k$, the genie decides to either reveal the indices of $k-1$ heavy balls or the indices of all $k$ heavy balls through a random process. Let $\ct\triangleq\{j \in \{0,1,\ldots,\alpha\}: \alpha(p-\eps)\le j \le \alpha(p+\eps)\}$ be the set of \emph{typical} bin indices for light balls. The genie picks a random bin $J \in \ct$ from distribution $\P(J=j)=\frac{L_j}{\sum_{i\in\ct}L_i}$.
If the chosen bin $J$ contains no heavy balls, i.e., $H_J=0$, then the genie reveals the indices of all $k$ heavy balls by setting
        $(\bar{\mathcal{H}}_{0},\ldots, \bar{\mathcal{H}}_{\alpha})=(\mathcal{H}_{0},\ldots, \mathcal{H}_{\alpha})\;\;\text{and}\;\;(\bar{\mathcal{L}}_{0},\ldots, \bar{\mathcal{L}}_{\alpha})=(\mathcal{L}_{0},\ldots, \mathcal{L}_{\alpha}).$
On the other hand, if $H_J\ge 1$, then the genie chooses a heavy ball $i$ in bin $J$ uniformly at random to be hidden, and reveals the indices of all other heavy balls. This is done by setting 
$(\bar{\mathcal{H}}_{0},\ldots, \bar{\mathcal{H}}_{\alpha})=(\mathcal{H}_{0},\ldots,\mathcal{H}_{J-1}, \mathcal{H}_{J}\setminus \{i\}, \mathcal{H}_{J+1}, \ldots, \mathcal{H}_{\alpha})$ and $(\bar{\mathcal{L}}_{0},\ldots, \bar{\mathcal{L}}_{\alpha})=(\mathcal{L}_{0},\ldots,\mathcal{L}_{J-1},\mathcal{L}_{J}\cup \{i\},\mathcal{L}_{J+1},\ldots,  \mathcal{L}_{\alpha}).$
\end{itemize}

It is not hard to see that with the additional noiseless information revealed by the genie, the performance of Algorithm $\mathcal{A}'$ does not deteriorate. Moreover, because in all cases the sets $\bar{\mathcal{H}}_0^\alpha$ only contain heavy balls, and the sets $\{\bar{\mathcal{L}_j}: j \notin \mathcal{T}\}$ only contain light balls, we assume without loss of generality that the adaptive phase of $\mathcal{A}'$ only reveals the weights of the balls in \{$\bar{\mathcal{L}_j}: j \in \mathcal{T}\}$. In the remaining proof, we analyze the error probability of $\mathcal{A}'$ in the presence of genie-aided information. 

\emph{\textbf{Why can we establish a tighter lower bound?}} Before sketching the proof of our lower bound with the genie-aided information, we first clarify the key distinctions between our analysis and that in \citet{feige1994computing}. We also highlight the main reason why our approach provides a tighter lower bound for the optimal query complexity. 

In the proof by~\cite{feige1994computing}, the error bound for $\mathcal{A}'$ is also established through a genie-aided analysis, where the genie picks the bin $J$ to hide a heavy ball from \emph{all possible bins} $\{0,\ldots,\alpha\}$ instead of choosing it from the set $\ct$. It is then shown that when $\alpha$ is small enough, every bin in $\{0,\ldots,\alpha\}$ contains at least one heavy ball with high probability, ensuring the genie hides a heavy ball in the chosen bin. Furthermore, it is shown that the noiseless queries in the adaptive phase would fail to find the hidden heavy ball with a constant probability, resulting in the scenario where the algorithm observes the $k-1$ heavy balls revealed by the genie, but cannot distinguish whether there is a $k$th hidden heavy ball. This leads to an overall error probability of at least $\delta$ for the algorithm.

In our analysis, we realize that having a heavy ball in every bin with high probability is not a necessary condition for establishing the desired error bound for Algorithm $\mathcal{A}'$. In our proof, the genie instead picks the bin $J$ from $\ct$ to hide a heavy ball, where $\ct$ is a typical set of bins for the light balls. We then focus on analyzing the statistical behavior of these bins in $\ct$. We show that for each $i\in\ct$, the probability that bin $i$ contains at least one heavy ball is $\omega(\delta)$. Moreover, conditioned on the event that a heavy ball is hidden by the genie, we further show that the information revealed by the genie and the noiseless queries in the adaptive phase are not informative enough for the algorithm to distinguish whether there is a $k$th heavy ball that is hidden.
This leads to the desired error bound. Notably, requiring each bin $i\in\ct$ to contain at least one heavy ball with probability $\omega(\delta)$ imposes a much less stringent condition on the choice of $\alpha$ comparing to requiring all the bins to contain at least one heavy ball with high probability. Consequently, the choice of $\alpha= \frac{(1-\epsilon)\log \frac{k}{\delta}}{\div(p-\epsilon\|1-p)}$ in our analysis is much higher than the choice of $\alpha$ in~\citet{feige1994computing}. Therefore, we can show that the error lower bound $\delta$ applies to an algorithm with higher query complexity. This leads to a tighter lower bound for the optimal query complexity.

\emph{\textbf{Error analysis for the optimal estimator.}}
Towards a contradiction, we assume that the worst-case error is at most $\delta$, i.e.,
\begin{equation}
\max\{\P(\widehat{\sfth}_k=1|w(\mathbf{x}) = k-1), \P(\widehat{\sfth}_k=0|w(\mathbf{x}) = k\} \le \delta.\label{eq:contradiction}
\end{equation}
This implies 
\begin{equation}
\label{eq:imply-from}
    \P(\widehat{\sfth}_k=0|w(\mathbf{x})=k-1)\ge 1-\delta.
\end{equation}
However, in the rest of the proof, we show~\eqref{eq:imply-from} implies
\begin{equation}
\label{eq:imply-to}
\P(\widehat{\sfth}_k=0|w(\mathbf{x})=k)> \delta,
\end{equation}
which contradicts with~\eqref{eq:contradiction}.

\emph{\textbf{Sufficient statistic for estimation.}}
We claim a sufficient statistic for the optimal estimator $\widehat{\sfth}_k$. When algorithm $\mathcal{A}'$ decides its output, it has access to information including $\bar{\mathcal{H}}_0^\alpha$, $\bar{\mathcal{L}}_0^\alpha$ and the weights of the balls revealed in the adaptive phase. By the symmetry of the balls in the same bin, it suffices for $\mathcal{A}'$ to consider the cardinality of the sets $\bar{\mathcal{H}}_0^\alpha$ and $\bar{\mathcal{L}}_0^\alpha$, denoted by $\bar{\bf H}_0^\alpha \triangleq (|\bar{\mathcal{H}}_0|,\ldots,|\bar{\mathcal{H}}_\alpha|)$ and $\bar{\bf L}_0^\alpha \triangleq (|\bar{\mathcal{L}}_0|,\ldots,|\bar{\mathcal{L}}_\alpha|)$. Moreover, notice that the unrevealed balls include at most $1$ heavy ball. The adaptive phase of the algorithm either finds one heavy ball or no heavy ball. By the symmetry of the balls in the same bin, it suffices for $\mathcal{A}'$ to consider whether the adaptive phase finds any heavy ball. Let $\ce_1$ denote the event that the adaptive phase does not find any heavy ball. Thus, the optimal estimator $\widehat{\sfth}_k$ of $\mathcal{A}'$ can be written as a function $\widehat{\sfth}_k(\bar{\bf H}_0^\alpha,\bar{\bf L}_0^\alpha,\mathbbm{1}_{\{\ce_1\}})$. 

Next, we argue that certain realizations of the sufficient statistic yield the correct answer to the true threshold $\sfth_k(\mathbf{x})$ (and thus no error for the optimal estimator). Notice that if $\sum_{i=0}^\alpha\bar{H}_i=k$, then we know that $w(\mathbf{x})=k$. This is because $\bar{\mathcal{H}}_0^\alpha$ only contain indices of heavy balls. Moreover, if $\mathbbm{1}_{\{\ce_1\}}=0$, we know that a heavy ball is found in the adaptive phase, and it follows that $w(\mathbf{x})=k$. Therefore, we assume without loss of generality that the optimal estimator $\widehat{\sfth}_k(\bar{\bf H}_0^\alpha,\bar{\bf L}_0^\alpha,\mathbbm{1}_{\{\ce_1\}})=1$ if $\sum_{i=0}^\alpha\bar{H}_i=k$ or $\mathbbm{1}_{\{\ce_1\}}=0$. 

Now the only ambiguous case that might lead to error is when $\sum_{i=1}^\alpha \bar{H}_i=k-1$ and $\mathbbm{1}_{\{\ce_1\}}=1$. This corresponds to the case where the genie reveals the indices of $k-1$ heavy balls, and the adaptive phase does not find any heavy balls. Among the remaining realizations of the sufficient statistic, we further divide them into two sets according to whether the estimator $\widehat{\sfth}_k({\bf h}_0^\alpha,{\bf l}_0^\alpha,1)$ is 0 or 1. We define $\mathcal{N}$ as the collection of realizations $({\bf h}_0^\alpha,{\bf l}_0^\alpha)$ such that $\widehat{\sfth}_k({\bf h}_0^\alpha,{\bf l}_0^\alpha,1)=0$.
In other words, we have 
\begin{align}
    \P(\widehat{\sfth}_k=0)
    =\sum_{({\bf h}_0^\alpha,{\bf l}_0^\alpha)\in \mathcal{N}}\P(\{(\bar{\bf H}_0^\alpha,\bar{\bf L}_0^\alpha)=({\bf h}_0^\alpha,{\bf l}_0^\alpha)\}\cap\ce_1).\label{eq:0-decomp}
\end{align}

Ideally, if we can prove that this sufficient statistic $(\bar{\bf H}_0^\alpha,\bar{\bf L}_0^\alpha,\mathbbm{1}_{\{\ce_1\}})$ is similarly distributed under the two different hypotheses $w(\bx)=k-1$ and $w(\bx)=k$, then we can lower bound the ratio $\frac{\P(\widehat{\sfth}_k=0|w(\bx)=k)}{\P(\widehat{\sfth}_k=0|w(\bx)=k-1)}$, and hence show that~\eqref{eq:imply-from} implies~\eqref{eq:imply-to}. More specifically, if we can show that 
\begin{equation}
\label{eq:ideal}
    \frac{\P\left(\{(\bar{\bf H}_0^\alpha,\bar{\bf L}_0^\alpha)=({\bf h}_0^\alpha,{\bf l}_0^\alpha)\}\cap\ce_1|w(\mathbf{x})=k\right)}{\P\left(\{(\bar{\bf H}_0^\alpha,\bar{\bf L}_0^\alpha)=({\bf h}_0^\alpha,{\bf l}_0^\alpha)\}\cap\ce_1|w(\mathbf{x})=k-1\right)}=\omega(\delta)
\end{equation}
for any $({\bf h}_0^\alpha,{\bf l}_0^\alpha)\in\mathcal{N}$, then we would have
\[
\P(\widehat{\sfth}_k=0|w(\bx)=k)=\omega(\delta)\P(\widehat{\sfth}_k=0|w(\bx)=k-1)>\delta,
\]
which completes the proof. However, it turns out that the desired property~\eqref{eq:ideal} does not hold for all realizations $({\bf h}_0^\alpha,{\bf l}_0^\alpha)\in\mathcal{N}$.
Instead, we slightly adjust this approach, and identify a family $\mathcal{C}$ of realizations $({\bf h}_0^\alpha,{\bf l}_0^\alpha)$ satisfying
\begin{enumerate}
    \item Typicality:
       $\sum_{({\bf h}_0^\alpha,{\bf l}_0^\alpha)\in\mathcal{C}}\P(\{(\bar{\bf H}_0^\alpha,\bar{\bf L}_0^\alpha)=({\bf h}_0^\alpha,{\bf l}_0^\alpha)\}|w(\mathbf{x})=k-1)\ge 1-o(1).$
    \item Bounded ratio property\footnote{For simplicity, the bounded ratio property presented here is a simplified version of what we actually prove, and it holds only in the special case when $\ca'$ makes a \emph{fixed} number of queries in the adaptive phase. We refer the readers to Propositions~\ref{prop:pmf-C} and~\ref{prop:pmf-D} for the properties in the general case.}: $\frac{\P\left(\{(\bar{\bf H}_0^\alpha,\bar{\bf L}_0^\alpha)=({\bf h}_0^\alpha,{\bf l}_0^\alpha)\}\cap\ce_1|w(\mathbf{x})=k\right)}{\P\left(\{(\bar{\bf H}_0^\alpha,\bar{\bf L}_0^\alpha)=({\bf h}_0^\alpha,{\bf l}_0^\alpha)\}\cap\ce_1|w(\mathbf{x})=k-1\right)}=\omega(\delta),$ $\forall ({\bf h}_0^\alpha,{\bf l}_0^\alpha)\in\mathcal{C}$.
\end{enumerate}
With the two properties for the family $\mathcal{C}$, we have
\begin{align*}
    \P(\widehat{\sfth}_k=0|w(\bx)=1)
    &\ge\sum_{({\bf h}_0^\alpha,{\bf l}_0^\alpha)\in \mathcal{N}\cap\mathcal{C}}\P(\{(\bar{\bf H}_0^\alpha,\bar{\bf L}_0^\alpha)=({\bf h}_0^\alpha,{\bf l}_0^\alpha)\}\cap\ce_1|w(\bx)=1)\\
    &\ge \omega(\delta)\sum_{({\bf h}_0^\alpha,{\bf l}_0^\alpha)\in \mathcal{N}\cap\mathcal{C}}\P(\{(\bar{\bf H}_0^\alpha,\bar{\bf L}_0^\alpha)=({\bf h}_0^\alpha,{\bf l}_0^\alpha)\}\cap\ce_1|w(\bx)=0)\\
    &\ge \omega(\delta(1-\delta-o(1)))=\omega(\delta),
\end{align*}
which completes the proof.
In the detailed proof presented in Appendix~\ref{appd:proof-conv}, we further separate the regime $\log k > \frac{\log(1/\delta)}{\log\log(1/\delta)}$ into two sub-regimes $\frac{\log(1/\delta)}{\log\log(1/\delta)}<\log k \le \log(1/\delta)\log\log(1/\delta)$ and $\log k >\log(1/\delta)\log\log(1/\delta)$. We present the detailed construction of the family $\mathcal{C}$ in these two sub-regimes in Appendix~\ref{appd:proof-conv}.

\section{Main Idea in the Proof of the Upper Bound }\label{sec:proof-upper}
In this section, we propose a noisy computing algorithm for the $\sfth_k$ function, and illustrate the main idea in the proof of Theorem~\ref{thm:threshold_achievability}.
We separately consider two regimes $k> \frac{n}{\log n}$ and $k\le\frac{n}{\log n}$.

\subsection{Regime \texorpdfstring{$k> \frac{n}{\log n}$}{Lg}}
To estimate a function of a binary sequence $\bx$, one natural idea is first to estimate each bit in $\bx$ and then to compute the function with the estimates. In this regime, the proposed algorithm exactly follows this idea.

The proposed algorithm estimates the bits in the sequence by employing an existing algorithm \textsc{ChechBit} as a subroutine. \textsc{CheckBit}~(Algorithm 1 in~\cite{gu2023optimal}) takes as input a single bit and a desired error probability $\delta$, and returns an estimate of the bit through repeated queries. A detailed description of the \textsc{CheckBit} algorithm is provided in Algorithm~\ref{alg:checkbit} in Appendix~\ref{appd:proof-ach}.
Lemma 13 in~\cite{gu2023optimal} shows that the algorithm makes at most $\frac{1}{1-2p}\left\lceil\frac{\log\frac{1-\delta}{\delta}}{\log\frac{1-p}{p}}\right\rceil$ queries in expectation and returns the correct value of the input bit with probability at least $1-\delta$. 

In the proposed algorithm, we call the \textsc{CheckBit} function to estimate $x_i$ for each $i\in [n]$ with error probability set to $\delta/n$, and obtain an estimate $\hat{x}_i$. The algorithm then outputs $\widehat{\sfth}_k(x)=\mathbbm{1}_{\{\sum_{i=1}^n\hat{x}_i\ge k\}}$ as the estimate. The pseudo-code of this algorithm is given in Algorithm~\ref{alg:checkeachbit}.

\begin{algorithm}[t]
    \LinesNumbered
    \DontPrintSemicolon
    \caption{Proposed \textsc{NoisyThreshold} algorithm for $k\ge \frac{n}{\log n}$}
    \label{alg:checkeachbit}
    \KwData{Bit sequence ${\bf x}\hspace*{-0.2em}=\hspace*{-0.2em}(x_1,\hspace*{-0.1em}\ldots\hspace*{-0.1em},x_n)$, parameter $k\leq \frac{n}{2}$, error probability $\delta$, noise probability $p$.}
    \KwResult{Estimate of $\sfth_k({\bf x})$.}
    $w \gets 0$\;
    \For{$i\in [n]$}{
        $\hat{x}_i\gets$ \textsc{CheckBit}($x_i$, $\delta/n$, $p$)\;
        $w\gets w+\hat{x}_i$\;
    }
    \Return $\mathbbm{1}_{\{\sum_{i=1}^n\hat{x}_i\ge k\}}$
\end{algorithm}

By the union bound, we have $\P(\exists i\in [n]: \hat{x}_i\neq x_i)\le \delta$. Consequently, we get $\P(\widehat{\sfth}_k(x)\neq \sfth_k(x))\le \delta$, i.e., the error probability of Algorithm~\ref{alg:checkeachbit} is at most $\delta$. The expected number of queries used by these $n$ calls of the \textsc{CheckBit} function is at most 
\[
\frac{n}{1-2p}\left\lceil\frac{\log\frac{1-\delta/n}{\delta/n}}{\log\frac{1-p}{p}}\right\rceil\le \frac{n}{1-2p}\left(\frac{\log\frac{n}{\delta}}{\log\frac{1-p}{p}}+1\right) =(1+o(1))\frac{n\log \frac{n}{\delta}}{\DKL}.
\]
In the regime $k>\frac{n}{\log n}$, it follows that $\log \frac{n}{\delta}=(1+o(1))\log \frac{k}{\delta}$. Therefore, Algorithm~\ref{alg:checkeachbit} uses at most $(1+o(1))\frac{n\log \frac{n}{\delta}}{\DKL}$ queries in expectation. This proves Theorem~\ref{thm:threshold_achievability} in the regime $k>\frac{n}{\log n}$.

Notably, in the opposite regime $k\le \frac{n}{\log n}$, the equation $\log \frac{n}{\delta}=(1+o(1))\log \frac{k}{\delta}$ does not necessarily hold true, and hence the desired upper bound in Theorem~\ref{thm:threshold_achievability} cannot be obtained through Algorithm~\ref{alg:checkeachbit}. For example, suppose $k=\log n$ and $\delta=1/\log n$. The upper bound provided by Algorithm~\ref{alg:checkeachbit} is $\frac{(1+o(1))n\log n}{D_\mathsf{KL}(p||1-p)}$, while the desired upper bound in Theorem~\ref{thm:threshold_achievability} is given by $\frac{(1+o(1))2n\log\log n}{D_\mathsf{KL}(p||1-p)}$.

\subsection{Regime \texorpdfstring{$k\le \frac{n}{\log n}$}{Lg}}
In this regime, we propose the \textsc{NoisyThreshold} algorithm, given in Algorithm~\ref{alg:proposed}, that computes the $\sfth_k$ function with the desired number of queries and error probability. 
In addition to \textsc{CheckBit}, Algorithm~\ref{alg:proposed} also utilizes the existing \textsc{MaxHeapThreshold} algorithm as a subroutine.

\setcounter{AlgoLine}{0}

\begin{algorithm}[t]
    \LinesNumbered
    \DontPrintSemicolon
    \caption{Proposed \textsc{NoisyThreshold} algorithm for $k\le \frac{n}{\log n}$}
    \label{alg:proposed}
    \KwData{Bit sequence ${\bf x}\hspace*{-0.2em}=\hspace*{-0.2em}(x_1,\hspace*{-0.1em}\ldots\hspace*{-0.1em},x_n)$, parameter $k\leq \frac{n}{2}$, error probability $\delta$, noise probability $p$.}
    \KwResult{Estimate of $\sfth_k({\bf x})$.}
    $\cs \gets \emptyset$\;
    
    \For{$i\in [n]$}{
        \If{\textsc{CheckBit}($x_i$, $\delta/k$, $p$) = $1$}{
            Append $x_i$ to $\cs$\;
        }
    }
    \uIf{$\abs{\cs} \leq k-1$}{
        \Return 0\;
    }
    \uElseIf{$\abs{\cs} \geq k+ \max(n\delta+n\sqrt{\delta},\frac{n}{\log n})$}{
        \Return 1\;
    }
    \Else{
        \Return \textsc{MaxHeapThreshold}($\cs$, $k$, $\delta$, $p$)\;
    }
\end{algorithm}

    
\textsc{MaxHeapThreshold} is the algorithm proposed by~\cite{feige1994computing} for estimating the $\sfth_k$ function. Its main idea is to use heapsort to find the $k$-th largest element in $\bx$, and then repeatedly query this element to decide $\sfth_k(x)$. To cope with the observation noise, each query in heapsort is repeated for a certain amount of times.
Detailed descriptions of the \textsc{MaxHeapThreshold} algorithm and its subroutines are given in Algorithms~\ref{alg:noisycomparereading},~\ref{alg:constructheap} and~\ref{alg:heap_threshold} in Appendix~\ref{appd:proof-ach}. 
Theorem 3.3 in~\cite{feige1994computing} shows that the function uses at most $O(n\log\frac{k}{\delta})$ queries. Although the query complexity of \textsc{MaxHeapThreshold} achieves the correct asymptotic order, there remains a constant multiplicative gap between its query complexity and the optimal value, as demonstrated in the numerical examples provided in Section~\ref{sec:introduction}.

The main idea in Algorithm~\ref{alg:proposed} is first reducing the problem size to the order of $o(n)$ through a filtering process for the bits in $\bx$, and then applying \textsc{MaxHeapThreshold} on the reduced problem. In the first step, we call the \textsc{CheckBit} function on each bit to estimate its value with tolerated error probability $\delta/k$. The bits estimated to be $1$ are added to the set $\mathcal{S}$. This step can be viewed as a filtering process that provides a set of bits believed to be $1$ with high confidence. Notice that with an error probability $\delta/k$ for each bit estimation, we cannot guarantee that all bits are correctly estimated with probability at least $1-\delta$ as in the case of Algorithm~\ref{alg:checkeachbit}. Instead, we show that with probability at least $1-2\delta$, $|\cs|$ is below a certain threshold $\tau$ if $\sfth_k(x)=0$, and $\cs$ contains at least $k$ bits of true value 1 if $\sfth_k(x)=1$. From here, we can output $\sfth_k(x)=1$ if $|\cs|\ge \tau$ (Line 10 of Algorithm~\ref{alg:proposed}), and output $\sfth_k(x)=0$ if $|\cs|\le k-1$ (Line 8 of Algorithm~\ref{alg:proposed}). When none of these two cases holds, the problem of estimating $\sfth_k(\bx)$ reduces to deciding whether $\cs$ contains at least $k$ bits of value $1$. For this reduced problem, we perform the second step by calling the \textsc{MaxHeapThreshold} function on the set $\cs$ (Line 12 of Algorithm~\ref{alg:proposed}) to decide the final output. 

By applying the union bound over the two steps, the overall error probability of Algorithm~\ref{alg:proposed} is at most $\delta'$ when the input parameter $\delta$ is set to $\delta/3$. Notably, $\tau$ is chosen to be $o(n)$. So in the event that the \textsc{MaxHeapThreshold} function is executed, its input size is $o(n)$. Although running \textsc{MaxHeapThreshold} induces a suboptimality of a constant factor, running it on an input size of $o(n)$ does not affect the dominating term in the query complexity of Algorithm~\ref{alg:proposed}. Therefore, the dominating term of the query complexity solely depends on the query complexity in the first step, which is upper bounded by $(1+o(1))\frac{n\log \frac{k}{\delta'}}{\DKL}$. This proves Theorem~\ref{thm:threshold_achievability} in the regime of $k\le\frac{n}{\log n}$. The detailed proof for this regime is presented in Appendix~\ref{appd:proof-ach}.

\section{Conclusion and Discussion}
In this work, we establish new upper and lower bounds of optimal query complexity for noisy computing of the $\sfth_k$ function. Our bounds are asymptotically tight in the case of $k=o(n)$, while having a multiplicative gap of at most $2$ in general. Our new bounds strictly improve the best known bounds by~\citet{feige1994computing}. In the following, we discuss several extensions of our study.

\paragraph{Closing the gap.} We believe the gap between our bounds stems from the lower bound. To establish the lower bound, we assume the existence of a genie that reveals auxiliary information to the algorithm, with the amount of information scaling with $k$. We conjecture that when $k$ is large, the additional information provided by the genie fundamentally alters the optimal query complexity, leading to a looser lower bound. How to tighten our lower bound remains an open problem.

\paragraph{Unknown $p$.} This paper assumes that the parameter $p$ is known. However, our bounds can be easily extended to the case of unknown $p$. Note that the noisy computing problem becomes strictly harder when $p$ is unknown, so our lower bound automatically holds in the unknown $p$ setting. To extend our upper bound, the proposed algorithms need to be modified so that they do not rely on the knowledge of $p$. The key idea is to incorporate an initial estimation step for $p$ before proceeding with the subsequent steps using the estimated value. More specifically, we can first query $x_1$ for $\theta=\frac{n\log\frac1\delta}{\log n}$ times, and compute a biased estimate of $p$ as $\hat{p}=\frac{\min(r,\theta-r)}{\theta(1-\frac{1}{\log n})}$, where $r$ denotes the number of $1$'s observed in these $\theta$ queries. 
The original steps in
Algorithms~\ref{alg:checkeachbit} and~\ref{alg:proposed} are then executed with input $\hat{p}$. By incorporating this estimation step, it can be shown that the error probabilities of Algorithms~\ref{alg:checkeachbit} and~\ref{alg:proposed} both increase by at most $o(\delta)$, and the expected query complexities both increase by a factor of at most $1+o(1)$. This extends our upper bound to the unknown $p$ setting.

\paragraph{Varying $p$.} In our study, we always assume that the noise parameter $p$ is a constant bounded away from both $0$ and $\frac12$. Another interesting extension of our study is to let $p$ also be a function of $n$, and considering the limiting behavior of the optimal complexity when $p\rightarrow 0$ or $p\rightarrow \frac12$.

\bibliographystyle{apalike}
\bibliography{ref.bib}

\newpage
\appendices
\section{Detailed Proof of Theorem~\ref{thm:thk-conv}}
\label{appd:proof-conv}
In this section, we prove Theorem~\ref{thm:thk-conv} in three separate regimes $\frac{\log(1/\delta)}{\log\log(1/\delta)}<\log k \le \log(1/\delta)\log\log(1/\delta)$, $\log k >\log(1/\delta)\log\log(1/\delta)$ and $\log k \le \frac{\log(1/\delta)}{\log\log(1/\delta)}$.
\subsection{Regime 1: \texorpdfstring{$\frac{\log(1/\delta)}{\log\log(1/\delta)}<\log k \le \log(1/\delta)\log\log(1/\delta)$}{Lg}}
\label{appd:proof-conv-Theta}

Let $\eps=(\log k)^{-1/4}$. Because we assume that $\log k>\frac{\log(1/\delta)}{\log\log(1/\delta)} =\omega(1)$, we have $\eps=o(1)$. Since the KL divergence is a Lipschitz continuous function, it follows that $\frac{(1-\epsilon)(n-k-\epsilon n)\log \frac{k}{\delta}}{\div(p-\epsilon\|1-p)}=(1-o(1))\frac{(n-k)\log\frac{k}{\delta}}{\DKL}$, and it suffices to show that for any algorithm that makes at most $\frac{(1-\epsilon)(n-k-\epsilon n)\log \frac{k}{\delta}}{\div(p-\epsilon\|1-p)}$ queries in expectation, the worst-case error probability is at least $\delta$.


Let $\mathcal{A}$ be an arbitrary algorithm that uses at most $\frac{(1-\epsilon)(n-k-\epsilon n)\log \frac{k}{\delta}}{\div(p-\epsilon\|1-p)}$ queries in expectation. Recall the enhanced algorithm $\mathcal{A}'$ described in Section~\ref{sec:proof-lower}. By Lemma~\ref{lem:enhancement}, it suffices to show that the worst case error probability of algorithm $\mathcal{A}'$ is at least $\delta$. To accomplish this, we prove 
\begin{equation}
\label{eq:imply-to-full}
\P(\widehat{\sfth}_k=0|w(\mathbf{x})=k)> \delta
\end{equation}
under the assumption that
\begin{equation}
\label{eq:imply-from-full}
    \P(\widehat{\sfth}_k=0|w(\mathbf{x})=k-1)\ge 1-\delta.
\end{equation}

Recall the notations under the balls-and-bins model we defined in Section~\ref{sec:proof-lower}. In addition, we define $P_j^1\triangleq\binom{\alpha}{j}(1-p)^j p^{\alpha-j}$ as the probability that a heavy ball is put into bin $j$, and $P_j^0\triangleq\binom{\alpha}{j}p^j (1-p)^{\alpha-j}$ as the probability that a light ball is put into bin $j$. Recall that we define $\mathcal{N}$ as the collection of realizations $({\bf h}_0^\alpha,{\bf l}_0^\alpha)$ such that $\widehat{\sfth}_k({\bf h}_0^\alpha,{\bf l}_0^\alpha,1)=0$, and we have 
\begin{align}
    \P(\widehat{\sfth}_k=0)
    =\sum_{({\bf h}_0^\alpha,{\bf l}_0^\alpha)\in \mathcal{N}}\P(\{(\bar{\bf H}_0^\alpha,\bar{\bf L}_0^\alpha)=({\bf h}_0^\alpha,{\bf l}_0^\alpha)\}\cap\ce_1).\label{eq:0-decomp-full}
\end{align}
In the following, we construct the family $\mathcal{C}$ of realizations such that the typicality and the bounded ratio property introduced in Section~\ref{sec:proof-lower} are satisfied. Let $\mathcal{C}$ be the collection of all $(2\alpha+2)$-tuples of natural numbers $({\bf h}_0^\alpha,{\bf l}_0^\alpha)$ satisfying
\begin{enumerate}
\setlength\itemsep{0.2em}
    \item $\sum_{i=0}^\alpha h_i=k-1$, $\sum_{i=0}^\alpha l_i=n-k+1$;
    
    \item $\sum_{i\in\ct}l_i\ge (n-k+1)\left(1-2\exp\left(-\frac{\sqrt{\log k}}{6p\div(p-\eps\|1-p)}\right)\right)\left(1-\sqrt{\frac{\log n}{n}}\right)$;
    
    \item $(1-\Delta_j^0)(n-k+1)P_j^0\le l_j\le (1+\Delta_j^0)(n-k+1)P_j^0$ for any $j\in\ct$, where $\Delta_j^0\triangleq\sqrt{\frac{\log n}{(n-k+1)P_j^0}}$;
    
    \item $h_i=0$ for any $i\in\ct$.
\end{enumerate}
Here we provide a brief interpretation of the last three conditions in the definition of $\mathcal{C}$. Firstly, as we have mentioned, $\ct$ represents the typical range of bins for the light balls. Therefore, condition 2 essentially states that only a vanishing fraction of the light balls will fall outside the typical bins. Secondly, as we will see in the later part of the proof, $P_j^0=\Omega(n^{-1+\Theta(1)})$ for each $j\in\ct$. This implies that $\Delta_j^0=o(1)$, and condition 3 requires that the number of light balls in each typical bin has a small deviation from its mean. Finally, we will later show that $\E[H_j]=o(1)$ for each $j\in\ct$, and condition 4 essentially requires the concentration of the number of heavy balls in these bins. From these, we can see that $\mathcal{C}$ is a collection of typical realizations for $({\bf H}_0^\alpha,{\bf L}_0^\alpha)$. In the following two propositions, we state the typicality and the bounded ratio property for $\mathcal{C}$.

\begin{proposition}
   \label{prop:prob-C} 
   There exists a positive constant $\gamma$ such that
   \begin{equation}
   \label{eq:prob-C}
       \sum_{({\bf h}_0^\alpha,{\bf l}_0^\alpha)\in\mathcal{C}}\P(\{({\bf H}_0^\alpha,{\bf L}_0^\alpha)=({\bf h}_0^\alpha,{\bf l}_0^\alpha)\}|w(\mathbf{x})=k-1)\ge 1-O\left(\frac{\log k\log\log(1/\delta)}{k^\frac{\gamma}{\log\log(1/\delta)}}\right)-O(n^{-1/6}).
   \end{equation}
\end{proposition}
\begin{proposition}
\label{prop:pmf-C}
    Let $n^*\triangleq(n-k+1)\left(1-2\exp\left(-\frac{\sqrt{\log k}}{6p\div(p-\eps\|1-p)}\right)\right)\left(1-\sqrt{\frac{\log n}{n}}\right)-(2\eps\alpha+1).$
    For any $({\bf h}_0^\alpha,{\bf l}_0^\alpha)\in\mathcal{C}$, we have 
    \begin{align}
        &\frac{\P\left(\{(\bar{\bf H}_0^\alpha,\bar{\bf L}_0^\alpha)=({\bf h}_0^\alpha,{\bf l}_0^\alpha)\}\cap\ce_1|w(\mathbf{x})=k\right)}{\P\left(\{(\bar{\bf H}_0^\alpha,\bar{\bf L}_0^\alpha)=({\bf h}_0^\alpha,{\bf l}_0^\alpha)\}|w(\mathbf{x})=k-1\right)}\label{eq:ratio}\\
        &\ge\tfrac{(1+o(1))\delta}{n}\left(\tfrac{k}{\delta}\right)^{\eps/2}\left(n^*-\E[N'|w(\bx)=k-1,(\bar{\bf H}_0^\alpha,\bar{\bf L}_0^\alpha)=({\bf h}_0^\alpha,{\bf l}_0^\alpha)]\right).\label{eq:pmf-C}
    \end{align}
\end{proposition}

With Propositions~\ref{prop:prob-C} and~\ref{prop:pmf-C}, we are now ready to prove inequality~\eqref{eq:imply-to-full}. By Proposition~\ref{prop:pmf-C}, we have
\begin{align}
    &\P(\widehat{\sfth}_k=0|w(\mathbf{x})=k)\nonumber\\
    &\ge \sum_{({\bf h}_0^\alpha,{\bf l}_0^\alpha)\in \mathcal{N}\cap\mathcal{C}}\P(\{(\bar{\bf H}_0^\alpha,\bar{\bf L}_0^\alpha)=({\bf h}_0^\alpha,{\bf l}_0^\alpha)\}\cap\ce_1|w(\mathbf{x})=k)\nonumber\\
    &\stackrel{(a)}{\ge}\sum_{({\bf h}_0^\alpha,{\bf l}_0^\alpha)\in \mathcal{N}\cap\mathcal{C}}\P(\{(\bar{\bf H}_0^\alpha,\bar{\bf L}_0^\alpha)=({\bf h}_0^\alpha,{\bf l}_0^\alpha)\}|w(\mathbf{x})=k-1)\nonumber\\
    &\hspace{1em}\cdot\tfrac{(1+o(1))\delta}{n}\left(\tfrac{k}{\delta}\right)^{\eps/2}\left(n^*-\E[N'|w(\bx)=k-1,(\bar{\bf H}_0^\alpha,\bar{\bf L}_0^\alpha)=({\bf h}_0^\alpha,{\bf l}_0^\alpha)]\right)\nonumber\\
    &=\tfrac{(1+o(1))\delta}{n}\left(\tfrac{k}{\delta}\right)^{\eps/2}\Bigg(\sum_{({\bf h}_0^\alpha,{\bf l}_0^\alpha)\in \mathcal{N}\cap\mathcal{C}}n^*\P(\{(\bar{\bf H}_0^\alpha,\bar{\bf L}_0^\alpha)=({\bf h}_0^\alpha,{\bf l}_0^\alpha)\}|w(\mathbf{x})=k-1)\nonumber\\
     &\hspace{1em} -\P(\{(\bar{\bf H}_0^\alpha,\bar{\bf L}_0^\alpha)=({\bf h}_0^\alpha,{\bf l}_0^\alpha)\}|w(\mathbf{x})=k-1)\E[N'|w(\bx)=k-1,(\bar{\bf H}_0^\alpha,\bar{\bf L}_0^\alpha)=({\bf h}_0^\alpha,{\bf l}_0^\alpha)]\Bigg).\label{eq:two_sums}
\end{align}
Now, we bound the summation in~\eqref{eq:two_sums}.
Since we assume that $\P(\widehat{\sfth}_k=0|w(\mathbf{x})=k-1)\ge 1-\delta$, we have
\begin{equation}
    1-\delta\le \P(\widehat{\sfth}_k=0|w(\mathbf{x})=k-1)
    =\sum_{({\bf h}_0^\alpha,{\bf l}_0^\alpha)\in \mathcal{N}}\P(\{(\bar{\bf H}_0^\alpha,\bar{\bf L}_0^\alpha)=({\bf h}_0^\alpha,{\bf l}_0^\alpha)\}|w(\mathbf{x})=k-1),\label{eq:th0-decomp}
\end{equation}
By Proposition~\ref{prop:prob-C}, 
we have
\[
\sum_{({\bf h}_0^\alpha,{\bf l}_0^\alpha)\in \mathcal{N}\cap\mathcal{C}}\P(\{(\bar{\bf H}_0^\alpha,\bar{\bf L}_0^\alpha)=({\bf h}_0^\alpha,{\bf l}_0^\alpha)\}|w(\mathbf{x})=k-1)\ge 1-\delta-O\left(\frac{\log k\log\log(1/\delta)}{k^\frac{\gamma}{\log\log(1/\delta)}}\right)-O(n^{-1/6}).
\]
By the definition of $n^*$ in Proposition~\ref{prop:pmf-C}, it follows that
\begin{align}
    &\sum_{({\bf h}_0^\alpha,{\bf l}_0^\alpha)\in \mathcal{N}\cap\mathcal{C}}n^*\P(\{(\bar{\bf H}_0^\alpha,\bar{\bf L}_0^\alpha)=({\bf h}_0^\alpha,{\bf l}_0^\alpha)\}|w(\mathbf{x})=k-1)\nonumber\\
    &\ge (n-k)\bigg(1-\delta-O\left(\frac{\log k\log\log(1/\delta)}{k^\frac{\gamma}{\log\log(1/\delta)}}\right)-O(n^{-1/6})-2\exp\left(-\frac{\sqrt{\log k}}{6p\div(p-\eps\|1-p)}\right)\nonumber\\
    &\;\;\;-O\left(\sqrt{\tfrac{\log n}{n}}\right)-O\left(\frac{\eps\alpha}{n}\right)\bigg).\label{eq:sum1}
\end{align}
By the law of total expectation and the fact that $N'$ is always non-negative, we have
\begin{align}
    &\sum_{({\bf h}_0^\alpha,{\bf l}_0^\alpha)\in \mathcal{N}\cap\mathcal{C}}\P(\{(\bar{\bf H}_0^\alpha,\bar{\bf L}_0^\alpha)=({\bf h}_0^\alpha,{\bf l}_0^\alpha)\}|w(\mathbf{x})=k-1)\E[N'|w(\bx)=k-1,(\bar{\bf H}_0^\alpha,\bar{\bf L}_0^\alpha)=({\bf h}_0^\alpha,{\bf l}_0^\alpha)]\nonumber\\
    &\le\E[N'|w(\bx)=k-1]=n-k-\eps n.\label{eq:sum2}
\end{align}
Because $\eps=(\log k)^{-1/4}$,
we have 
\[
\delta+\frac{\log k\log\log(1/\delta)}{k^\frac{\gamma}{\log\log(1/\delta)}}+n^{-1/6}+2\exp\left(-\frac{\sqrt{\log k}}{6p\div(p-\eps\|1-p)}\right)+\sqrt{\tfrac{\log n}{n}}+\frac{\eps\alpha}{n}=o(\epsilon).
\]
Plugging \eqref{eq:sum1} and \eqref{eq:sum2} into~\eqref{eq:two_sums} yields $$\P(\widehat{\sfth}_k=0|w(\mathbf{x})=k)\ge (1-o(1))\eps\delta\left(\tfrac{k}{
    \delta
    }\right)^{\eps/2}.$$
    Finally, $\eps= (\log k)^{-1/4}$ implies that $\eps\left(\tfrac{k}{\delta}\right)^{\eps/2}=\omega(1)$. This shows that $$\P(\widehat{\sfth}_k=0|w(\mathbf{x})=k)\ge \omega(\delta).$$ 
    
To finish the proof in this regime, we are left to prove Propositions~\ref{prop:prob-C} and~\ref{prop:pmf-C} and Lemma~\ref{lem:enhancement}.

\subsubsection{Proof of Propositions}
\renewcommand*{\proofname}{Proof of Proposition~\ref{prop:prob-C}}
    \begin{proof}
        We define error events 
        $$\ce_2\triangleq\left\{\sum_{j\in\ct}L_j\le (n-k+1)\left(1-2\exp\left(-\frac{\sqrt{\log k}}{6p\div(p-\eps\|1-p)}\right)\right)\left(1-\sqrt{\frac{\log n}{n}}\right)\right\},$$
        $$\ce_3\triangleq\{\exists j\in\ct:L_j\notin[(1-\Delta_j^0)(n-k+1)P_j^0,(1+\Delta_j^0)(n-k+1)P_j^0]\},$$
        and
        $$\ce_4\triangleq\{\exists j\in\ct:H_j\ge 1\}.$$
        Notice that events $\ce_2,\ce_3$ and $\ce_4$ correspond to the second, third and fourth conditions in the definition of $\mathcal{C}$. The first condition is always satisfied because we always have $\sum_{i=0}^\alpha H_i=k-1$ and $\sum_{i=0}^\alpha L_i=n-k+1$ given $w(\mathbf{x})=k-1$.
        It follows that
        \begin{align*}
            &\sum_{({\bf h}_0^\alpha,{\bf l}_0^\alpha)\in\mathcal{C}}\P(\{({\bf H}_0^\alpha,{\bf L}_0^\alpha)=({\bf h}_0^\alpha,{\bf l}_0^\alpha)\}|w(\mathbf{x})=k-1)\\
            &= 1-\P(\ce_2\cup\ce_3\cup\ce_4|w(\mathbf{x})=k-1)\\
            &\ge 1-\P(\ce_2|w(\mathbf{x})=k-1)-\P(\ce_3|w(\mathbf{x})=k-1)-\P(\ce_4|w(\mathbf{x})=k-1),
        \end{align*}
        where the last inequality follows by the union bound. We bound the probabilities of these three error events in the following lemmas. The proof of these lemmas are deferred to later part of this section.
        \begin{lemma}
        \label{lem:ce2}
        For any $\mathbf{x}$ with $w(\mathbf{x})=k-1$, we have
        $\P(\ce_2|\mathbf{x})\le n^{-1/5}.$
        \end{lemma}
        \begin{lemma}
    \label{lem:ce3}
        For any $\mathbf{x}$ with $ w(\mathbf{x})=k-1$, we have $\P(\ce_3|\mathbf{x})\le n^{-1/6}$.
    \end{lemma}
    \begin{lemma}
    \label{lem:ce4}
        For any $\mathbf{x}$ with $ w(\mathbf{x})=k-1$, we have $\P(\ce_4|\mathbf{x})=O\left(\frac{\log k\log\log(1/\delta)}{k^\frac{\gamma}{\log\log(1/\delta)}}\right)$, where $\gamma$ is a positive constant.
    \end{lemma}
    By Lemmas \ref{lem:ce2}, \ref{lem:ce3} and \ref{lem:ce4}, we have
    \begin{align*}
        \sum_{({\bf h}_0^\alpha,{\bf l}_0^\alpha)\in\mathcal{C}}\P(\{(\bar{\bf H}_0^\alpha,\bar{\bf L}_0^\alpha)=({\bf h}_0^\alpha,{\bf l}_0^\alpha)\}|w(\mathbf{x})=k-1)&\ge 1-n^{-1/5}-n^{-1/6}-O\left(\frac{\log k\log\log(1/\delta)}{k^\frac{\gamma}{\log\log(1/\delta)}}\right)\\
        &=1-O(n^{-1/6})-O\left(\frac{\log k\log\log(1/\delta)}{k^\frac{\gamma}{\log\log(1/\delta)}}\right),
    \end{align*}
    which completes the proof of Proposition~\ref{prop:prob-C}.
    \end{proof}

    \renewcommand*{\proofname}{Proof of Proposition~\ref{prop:pmf-C}}
\begin{proof}
    To begin with, we notice that on any input $\mathbf{x}$ with $w(\mathbf{x})=k-1$, the genie-aided information is exactly $\bar{\bf H}_0^\alpha={\bf H}_0^\alpha$ and $\bar{\bf L}_0^\alpha={\bf L}_0^\alpha$. Moreover, conditioned on $w(\mathbf{x})=k-1$, we know that ${\bf H}_0^\alpha$ represents the number of heavy balls in each bin, and ${\bf L}_0^\alpha$ represents the number of light balls in each bin. Therefore, given $w(\mathbf{x})=k-1$, ${\bf H}_0^\alpha$ and ${\bf L}_0^\alpha$ are two independent random vectors with multinomial conditional distributions ${\bf H}_0^\alpha|\{w(\mathbf{x})=k-1\}\sim\mathrm{M}(k-1;P_0^1,\ldots,P_\alpha^1)$ and ${\bf L}_0^\alpha|\{w(\mathbf{x})=k-1\}\sim\mathrm{M}(n-k+1;P_0^0,\ldots,P_\alpha^0)$ respectively. Here, we say a random vector $\mathbf{Z} = (Z_1,\ldots,Z_m)$ follows multinomial distribution $\mathrm{M}(N;p_1,\ldots,p_m)$ if its pmf is given by $\P(\mathbf{Z}=\mathbf{z})=\frac{N!\prod_{i=1}^m p_i^{z_i}}{\prod_{i=1}^m z_i!}.$ Therefore, the denominator in \eqref{eq:ratio} can be written as
    \begin{align}
        &\P\left(\{(\bar{\bf H}_0^\alpha,\bar{\bf L}_0^\alpha)=({\bf h}_0^\alpha,{\bf l}_0^\alpha)\}|w(\mathbf{x})=k-1\right)\nonumber\\
        &=\frac{(k-1)!\prod_{i=0}^\alpha (P_i^1)^{h_i}}{\prod_{i=0}^\alpha h_i!}\frac{(n-k+1)!\prod_{i=0}^\alpha (P_i^0)^{l_i}}{\prod_{i=0}^\alpha l_i!}.\label{eq:rhs-prob}
    \end{align}
    Now, we move on to consider the numerator in~\eqref{eq:ratio}. Recalled that when $w(\mathbf{x})=k$, the genie picks a bin with random index $J\in\ct$. If there exists a heavy ball in bin $J$, the genie randomly chooses a heavy ball from there and keeps it unrevealed. The numerator in \eqref{eq:ratio} can be written as
    \begin{align}
        &\P\left(\{(\bar{\bf H}_0^\alpha,\bar{\bf L}_0^\alpha)=({\bf h}_0^\alpha,{\bf l}_0^\alpha)\}\cap\ce_1|w(\mathbf{x})=k\right)\nonumber\\
        &=\sum_{j\in\ct}\P\left(\{(\bar{\bf H}_0^\alpha,\bar{\bf L}_0^\alpha)=({\bf h}_0^\alpha,{\bf l}_0^\alpha)\}\cap\{J=j\}\cap\ce_1|w(\mathbf{x})=k\right)\nonumber\\
        &=\sum_{j\in\ct}\P\left(\{({\bf H}_0^\alpha,{\bf L}_0^\alpha)=({\bf h}_0^\alpha+{\bf e}_j,{\bf l}_0^\alpha-{\bf e}_j)\}\cap\{J=j\}\cap\ce_1|w(\mathbf{x})=k\right),\label{eq:decomp-J}
    \end{align}
    where for each $j\in\{0,\ldots,\alpha\}$, ${\bf e}_j$ is the length-$(\alpha+1)$ indicator vector for the $(j+1)$-th coordinate, i.e., the $(j+1)$-th coordinate has value one, while all other coordinates have value zero. By the chain rule, we have
    \begin{align}
        &\P\left(\{({\bf H}_0^\alpha,{\bf L}_0^\alpha)=({\bf h}_0^\alpha+{\bf e}_j,{\bf l}_0^\alpha-{\bf e}_j)\}\cap\{J=j\}\cap\ce_1|w(\mathbf{x})=k\right)\nonumber\\
        &=\P(\{{\bf H}_0^\alpha={\bf h}_0^\alpha+{\bf e}_j\}\cap\{{\bf L}_0^\alpha={\bf l}_0^\alpha-{\bf e}_j\}|w(\mathbf{x})=k)\nonumber\\
        &\;\;\cdot\P(\{J=j\}|{\bf H}_0^\alpha={\bf h}_0^\alpha+{\bf e}_j,{\bf L}_0^\alpha={\bf l}_0^\alpha-{\bf e}_j,w(\mathbf{x})=k)\nonumber\\
        &\;\;\cdot\P(\ce_1|J=j,Y_0^\alpha=y_0^\alpha+{\bf e}_j,Z_0^\alpha=z_0^\alpha-{\bf e}_j,w(\mathbf{x})=k),\label{eq:k-chain}
    \end{align}
    Now, we bound the three probabilities in~\eqref{eq:k-chain} separately.
    Firstly, given $w(\mathbf{x})=k$, ${\bf H}_0^\alpha$ and ${\bf L}_0^\alpha$ are two independent random vectors with conditional distributions ${\bf H}_0^\alpha|\{w(\mathbf{x})=k\}\sim\mathrm{M}(k;P_0^1\ldots,P_\alpha^1)$ and ${\bf L}_0^\alpha|\{w(\mathbf{x})=k\}\sim\mathrm{M}(n-k;P_0^0\ldots,P_\alpha^0)$ respectively. This suggests that
    \begin{align}
        \P(\{{\bf H}_0^\alpha={\bf h}_0^\alpha+{\bf e}_j\}\cap\{{\bf L}_0^\alpha={\bf l}_0^\alpha-{\bf e}_j\}|w(\mathbf{x})=k)
        =\frac{k!P_j^1\prod_{i=0}^\alpha (P_i^1)^{h_i}}{(h_j+1)\prod_{i=0}^\alpha h_i!}\frac{(n-k)!l_j\prod_{i=0}^\alpha (P_i^0)^{l_i}}{P_j^0\prod_{i=0}^\alpha l_i!}.\label{eq:term1}
    \end{align}
    Secondly, 
    \begin{align}
        \P(\{J=j\}|{\bf H}_0^\alpha={\bf h}_0^\alpha+{\bf e}_j,{\bf L}_0^\alpha={\bf l}_0^\alpha-{\bf e}_j,w(\mathbf{x})=k)=\frac{l_j-1}{-1+\sum_{j\in\ct}l_j}\label{eq:term2}
    \end{align}
    by the random process of choosing $J$.
    Finally, by the symmetry among the balls in the same bin, it suffices for the adaptive phase of the algorithm to choose the number of balls to be revealed in each bin. 
    Recall that we assume without loss of generality that the adaptive phase only reveals balls in sets $\{\bar{\mathcal{L}}_j: j\in \mathcal{T}\}$. 
    For $j\in\mathcal{T}$, let $N'_j$ denote the number of balls revealed in $\bar{\mathcal{L}}_j$. Note that 
    $N'_j$'s are random variables satisfying $\sum_{j\in\mathcal{T}}N'_j=N'$ by the construction of $\mathcal{A}'$. Notice that $J$ and $w(\bx)$ are both variables hidden from $\mathcal{A}'$. Moreover, the adaptive phase can only lead to two possible outcomes: a heavy ball is found or no heavy ball is found. Therefore, we can assume without loss of generality that the distribution of $N'_j$'s
    only depends on the observations $\bar{\mathcal{H}}_0^\alpha$ and $\bar{\mathcal{L}}_0^\alpha$. Because the probability of finding the heavy ball in $\bar{\mathcal{L}}_j$ by revealing $i$ balls is $\frac{i}{l_j}$, it follows that
    \begin{align}
        &\P(\ce_1|J=j,{\bf H}_0^\alpha={\bf h}_0^\alpha+{\bf e}_j,{\bf L}_0^\alpha={\bf l}_0^\alpha-{\bf e}_j,w(\mathbf{x})=k)\nonumber\\
        &=\sum_{i=0}^{l_j}\P(N'_j=i|J=j,{\bf H}_0^\alpha={\bf h}_0^\alpha+{\bf e}_j,{\bf L}_0^\alpha={\bf l}_0^\alpha-{\bf e}_j,w(\mathbf{x})=k)\left(1-\frac{i}{l_j}\right)\nonumber\\
        &\ge 1-\frac{\E[N'_j|J=j,{\bf H}_0^\alpha={\bf h}_0^\alpha+{\bf e}_j,{\bf L}_0^\alpha={\bf l}_0^\alpha-{\bf e}_j,w(\mathbf{x})=k]}{l_j-1}\nonumber\\
        &= 1-\frac{\E[N'_j|J=j,\bar{\bf H}_0^\alpha={\bf h}_0^\alpha,\bar{\bf L}_0^\alpha={\bf l}_0^\alpha,w(\mathbf{x})=k]}{l_j-1}\nonumber\\
        &=1-\frac{\E[N'_j|\bar{\bf H}_0^\alpha={\bf h}_0^\alpha,\bar{\bf L}_0^\alpha={\bf l}_0^\alpha,w(\mathbf{x})=k-1]}{l_j-1},\label{eq:term3}
    \end{align}
    where the penultimate equality follows because $\{J=j,{\bf H}_0^\alpha={\bf h}_0^\alpha+{\bf e}_j,{\bf L}_0^\alpha={\bf l}_0^\alpha-{\bf e}_j\}=\{J=j,\bar{\bf H}_0^\alpha={\bf h}_0^\alpha,\bar{\bf L}_0^\alpha={\bf l}_0^\alpha\}$, and the last equality follows because $N'_j$ is conditionally independent of $w(\bx)$ and $J$ given $\bar{\bf H}_0^\alpha$ and $\bar{\bf L}_0^\alpha$. Plugging \eqref{eq:term1}, \eqref{eq:term2} and \eqref{eq:term3} into \eqref{eq:k-chain} yields
        \begin{align}
        &\P\left(\{({\bf H}_0^\alpha,{\bf L}_0^\alpha)=({\bf h}_0^\alpha+{\bf e}_j,{\bf l}_0^\alpha-{\bf e}_j)\}\cap\{J=j\}\cap\ce_1|w(\mathbf{x})=k\right)\nonumber\\
        &\ge \frac{k!P_j^1\prod_{i=0}^\alpha (P_i^1)^{h_i}}{(h_j+1)\prod_{i=0}^\alpha h_i!}\cdot\frac{(n-k)!l_j\prod_{i=0}^\alpha (P_i^0)^{l_i}}{P_j^0\prod_{i=0}^\alpha l_i!}\cdot\frac{l_j-1-\E[N'_j|\bar{\bf H}_0^\alpha={\bf h}_0^\alpha,\bar{\bf L}_0^\alpha={\bf l}_0^\alpha,w(\mathbf{x})=k-1]}{-1+\sum_{j\in\ct} l_j}.\label{eq:together}
    \end{align}
    Then, we plug \eqref{eq:together} into \eqref{eq:decomp-J} and divide by \eqref{eq:rhs-prob} to get
    \begin{align}
        &\frac{\P\left(\{(\bar{\bf H}_0^\alpha,\bar{\bf L}_0^\alpha)=({\bf h}_0^\alpha,{\bf l}_0^\alpha)\}\cap\ce_1|w(\mathbf{x})=k\right)}{\P\left(\{(\bar{\bf H}_0^\alpha,\bar{\bf L}_0^\alpha)=({\bf h}_0^\alpha,{\bf l}_0^\alpha)\}|w(\mathbf{x})=k-1\right)}\nonumber\\
        &=\sum_{j\in\ct}\frac{kP_j^1}{h_j+1}\cdot\frac{l_j}{(n-k+1)P_j^0}\cdot\frac{l_j-1-\E[N'_j|\bar{\bf H}_0^\alpha={\bf h}_0^\alpha,\bar{\bf L}_0^\alpha={\bf l}_0^\alpha,w(\mathbf{x})=k-1]}{-1+\sum_{i\in\ct} l_i}.\label{eq:pmf-ratio}
    \end{align}%
    To complete the proof, we need a technical lemma that bounds the probabilities $P_j^1$ and $P_j^0$. 
    \begin{lemma}
    \label{lem:P1-0}
        For each $j\in\ct$, we have $P_j^0\ge \exp(-o(\log k))$, and $\frac{\delta}{k}\left(\frac{k}{\delta}\right)^{\eps/2}\le P_j^1 \le k^{-1-\frac{\gamma}{\log\log(1/\delta)}}$ for some positive constant $\gamma$.
    \end{lemma}
    Because $({\bf h}_0^\alpha,{\bf l}_0^\alpha)\in\mathcal{C}$, we have $h_j=0$ and $(1-\Delta_j^0)(n-k+1)P_j^0\le l_i\le (1+\Delta_j^0)(n-k+1)P_j^0$ for each $j\in\ct$. Recall that we defined $\Delta_j^0=\sqrt{\frac{\log n}{(n-k+1)P_j^0}}$. It follows by Lemma~\ref{lem:P1-0} that $\Delta_j^0=o(1)$ for each $j\in\ct$, which further implies that $l_j=(1+o(1))(n-k+1)P_j^0$. We then have
    \begin{align*}
        &\sum_{j\in\ct}\frac{kP_j^1}{h_j+1}\cdot\frac{l_j}{(n-k+1)P_j^0}\cdot\frac{l_j-1-\E[N'_j|\bar{\bf H}_0^\alpha={\bf h}_0^\alpha,\bar{\bf L}_0^\alpha={\bf l}_0^\alpha,w(\mathbf{x})=k-1]}{-1+\sum_{i\in\ct} l_i}\\
        &\ge\delta\left(\frac{k}{\delta}\right)^{\eps/2}(1+o(1))\sum_{j\in\ct}\frac{l_j-1-\E[N'_j|\bar{\bf H}_0^\alpha={\bf h}_0^\alpha,\bar{\bf L}_0^\alpha={\bf l}_0^\alpha,w(\mathbf{x})=k-1]}{-1+\sum_{i\in\ct} l_i}\\
        &\stackrel{(a)}{\ge} \frac{\delta}{n}\left(\frac{k}{\delta}\right)^{\eps/2}(1+o(1))\left(-(2\eps\alpha+1)-\E[N'|\bar{\bf H}_0^\alpha={\bf h}_0^\alpha,\bar{\bf L}_0^\alpha={\bf l}_0^\alpha,w(\mathbf{x})=k-1]+\sum_{j\in\ct}l_j\right)\\
        &\stackrel{(b)}{\ge} \frac{(1+o(1))\delta}{n}\left(\frac{k}{\delta}\right)^{\eps/2}\left(n^*-\E[N'|w(\bx)=k-1,(\bar{\bf H}_0^\alpha,\bar{\bf L}_0^\alpha)=({\bf h}_0^\alpha,{\bf l}_0^\alpha)]\right),
    \end{align*}
    where (a) follows because $-1+\sum_{i\in\ct}l_i\le n$ and (b) follows because $$\sum_{i\in\ct}l_i\ge (n-k+1)\left(1-2\exp\left(-\frac{\sqrt{\log k}}{6p\div(p-\eps\|1-p)}\right)\right)\left(1-\sqrt{\frac{\log n}{n}}\right)$$ on the assumption that $({\bf h}_0^\alpha,{\bf l}_0^\alpha)\in\mathcal{C}$.
\end{proof}

    \subsubsection{Proof of Lemmas}
    \renewcommand*{\proofname}{Proof of Lemma~\ref{lem:enhancement}}
\begin{proof}
    The key idea in this proof is to show that the original Algorithm $\mathcal{A}$ can be simulated by executing Algorithm $\mathcal{A}'$. 

    To show this simulation, we assume that the non-adaptive phase of $\mathcal{A}'$ is completed, and simulate the execution of $\mathcal{A}$ in the adaptive phase of $\mathcal{A}'$.
    When Algorithm $\mathcal{A}$ makes a query on bit $x_j$ for the $i$th time, where $i\le \alpha$, Algorithm $\mathcal{A}'$ returns the response to the $i$th query on $x_j$ from its non-adaptive phase to Algorithm $\mathcal{A}$. When Algorithm $\mathcal{A}$ makes a query on bit $x_j$ for the $\alpha+1$st time, Algorithm $\mathcal{A}'$ noiselessly reads the bit $x_j$, and returns response $x_j\oplus \mathsf{Bern}(p)$ to $\mathcal{A}$. When Algorithm $\mathcal{A}$ makes a query on bit $x_j$ for the $i$th time, where $i\ge \alpha+2$, because Algorithm $\mathcal{A}'$ already has the value of $x_j$, it returns response $x_j\oplus \mathsf{Bern}(p)$ to Algorithm $\mathcal{A}$. 
    Let $N$ denote the number of bits that are queried by $\mathcal{A}$ for at least $\alpha$ times. Recall that $M$ denotes the total number of queries made by $\mathcal{A}$.  Because we assume $\E[M|\mathbf{x}]\le \frac{(1-\epsilon)(n-k-\epsilon n)\log \frac{k}{\delta}}{\div(p-\epsilon\|1-p)}$, we have
    \begin{equation*}
        \frac{(1-\epsilon)(n-k-\epsilon n)\log \frac{k}{\delta}}{\div(p-\epsilon\|1-p)}\ge \E[M|\mathbf{x}]\ge\sum_{i=0}^n\P(N=i|\mathbf{x})\cdot i\alpha =\alpha\E[N|\mathbf{x}], 
    \end{equation*}
    for any input instance $\bx$, and it follows that $\E[N|\mathbf{x}]\le n-k-\eps n$.
    Thus, this simulation process uses at most $n-k-\eps n$ noiseless bits read in expectation in the adaptive phase of Algorithm $\mathcal{A}'$. Finally, it is not hard to see that the output distribution of this simulation is the same as the output distribution of Algorithm $\mathcal{A}$. 
\end{proof}
    \renewcommand*{\proofname}{Proof of Lemma~\ref{lem:P1-0}}
    \begin{proof}
        By definition, we have $P_j^1=\binom{\alpha}{j}(1-p)^jp^{\alpha-j}$. Notice that $\min_{j\in\ct}P_j^1=P_{\alpha(p-\eps)}^1$. By Stirling's approximation, we have
        \begin{align*}
            P_{\alpha(p-\eps)}^1&=\binom{\alpha}{(p-\eps)\alpha}(1-p)^{\alpha(p-\eps)}p^{\alpha(1-p+\eps)}\\
            &\ge \exp\left(-\alpha\left(\div(p-\eps\|1-p)+O\left(\frac{\log\alpha}{\alpha}\right)\right)\right)\\
            &=\exp\left(-\frac{(1-\eps)\log\frac{k}{\delta}}{\div(p-\eps\|1-p)}\left(\div(p-\eps\|1-p)+O\left(\frac{\log\alpha}{\alpha}\right)\right)\right)\\
            &=\left(\frac{\delta}{k}\right)^{(1-\eps)(1+O((\log\alpha)/\alpha))}\\
            &\ge \frac{\delta}{k}\left(\frac{k}{\delta}\right)^{\eps/2}
        \end{align*}
        To see the upper bounds $P_j^1\le k^{-1-\gamma}$, we observe that $\max_{j\in\ct}P_j^1=P_{\alpha(p+\eps)}^1$. By Stirling's approximation, we have
        \begin{align*}
            P_{\alpha(p+\eps)}^1&=\binom{\alpha}{(p+\eps)\alpha}(1-p)^{\alpha(p+\eps)}p^{\alpha(1-p-\eps)}\\
            &\le \exp(-\alpha\div(p+\eps\|1-p))\\
            &=\exp\left(-\frac{(1-\eps)\log\frac{k}{\delta}}{\div(p-\eps\|1-p)}\div(p+\eps\|1-p)\right)\\
            &\stackrel{(a)}{=}\exp\left(-(1-\eps)(1-O(\eps))\log\frac{k}{\delta}\right)\\
            &\le k^{-1-\frac{\gamma}{\log\log(1/\delta)}},
        \end{align*}
        where (a) follows because $\frac{\div(p+\eps\|1-p)}{\div(p-\eps\|1-p)}=1-O(\eps)$, and the existence of positive constant $\gamma$ is guaranteed by the fact that $\log\frac{1}{\delta}\ge\frac{\log k}{\log\log(1/\delta)}$ and $\eps=o(\frac{1}{\log\log(1/\delta)})$.

        We have defined $P_j^0=\binom{\alpha}{j}p^j(1-p)^{\alpha-j}$. We can see that $\min_{j\in\ct}P_j^0=\min(P_{\alpha(p-\eps)}^0,P_{\alpha(p+\eps)}^0)$. Therefore, it suffices to show that $P_{\alpha(p-\eps)}^0\ge \exp(-o(\log k))$ and $P_{\alpha(p+\eps)}^0\ge \exp(-o(\log k))$. By Stirling's approximation, we have
        \begin{align*}
            P_{\alpha(p-\eps)}^0&=\binom{\alpha}{\alpha(p-\eps)}p^{\alpha(p-\eps)}(1-p)^{\alpha(1-p+\eps)}\\
            &\ge \exp\left(-\alpha\left(\div(p-\eps\|p)+O\left(\frac{\log\alpha}{\alpha}\right)\right)\right)\\
            &=\exp\left(-\frac{(1-\eps)\log\frac{k}{\delta}}{\div(p-\eps\|1-p)}\left(\div(p-\eps\|p)+O\left(\frac{\log\alpha}{\alpha}\right)\right)\right)\\
            &\ge \exp(-o(\log k)),
        \end{align*}
        where the last inequality follows because $\div(p-\eps\|p)=O(\epsilon^2)$ and $\log\frac{k}{\delta}\le \log k+\log\log(1/\delta)\log k$.
        It can be similarly shown that $P_{\alpha(p+\eps)}^0\ge \exp(-o(\log k))$.
    \end{proof}
    \renewcommand*{\proofname}{Proof of Lemma~\ref{lem:ce2}}    
    \begin{proof}
        Let $W_0\sim\Binom(\alpha,p)$. For any $\mathbf{x}$ with $w(\mathbf{x})=k-1$, we have $\sum_{j\in\ct}L_j|\mathbf{x}\sim\Binom(n-k+1,\P(W_0\in\ct))$. To bound $\P(\ce_2|\mathbf{x})$, we first bound $\P(W_0\in\ct)$. We have 
\begin{align*}
    \P(W_0\notin\ct)&=\P(|W_0-\alpha p|\ge \alpha\eps)\\
    &\stackrel{(a)}{\le}2\exp\left(-\frac{\eps^2\alpha}{3p}\right)\\
    &\stackrel{(b)}{\le}2\exp\left(-\frac{(1-\eps)\log\frac{k}{\delta}}{3p\div(p-\eps\|1-p)\sqrt{\log k}}\right)\\
    &\stackrel{(c)}{\le} 2\exp\left(-\frac{\sqrt{\log k}}{6p\div(p-\eps\|1-p)}\right),
\end{align*}
where (a) follows by the Chernoff bound (see, for example, Corollary 4.6 in~\cite{mitzenmacher2017probability}), (b) follows because $\eps= (\log k)^{-1/4}$, and (c) follows because $\frac12(1-\eps)\log\frac{k}{\delta}\ge \log k$. For simplicity, we denote $c=\frac{1}{6p\div(p-\eps\|1-p)}$.
    It follows that
    \begin{align*}
        &\P(\ce_2|\mathbf{x})\\
        &\le \P\Bigg(\mathrm{Binom}\left(n-k+1,1-2\exp\left(-c\sqrt{\log k}\right)\right)\le\\
        &\hspace{2em}(n-k+1)\left(1-2\exp\left(-c\sqrt{\log k}\right)\right)\left(1-\sqrt{\frac{\log n}{n}}\right)\Bigg)\\
        &\stackrel{(d)}{\le} \exp\left(-\frac{(n-k+1)\left(1-2\exp(-c\sqrt{\log k})\right)\log n}{2n}\right)\\
        &\le n^{-1/5},
    \end{align*}
    where (d) follows by the Chernoff bound.
    \end{proof}
    \renewcommand*{\proofname}{Proof of Lemma~\ref{lem:ce3}}  
    \begin{proof}
        For any $\mathbf{x}$ with $w(\mathbf{x})=k-1$, we have $L_j|\mathbf{x}\sim\Binom(n-k+1,P_j^0)$ for each $j\in\ct$. For any $j\in\ct$, by Chernoff bound we have
        \begin{align*}
            &\P(|L_j-(n-k+1)P_j^0|\ge \Delta_j^0(n-k+1)P_j^0|\mathbf{x})\\
            &\le 2\exp(-(\Delta_j^0)^2(n-k+1)P_j^0/3)\\
            &=2n^{-1/3},
        \end{align*}
    where the last equality follows because $\Delta_j^0=\sqrt{\frac{\log n}{(n-k+1)P_j^0}}$. By union bound, we have
        \begin{equation*}
            \P(\ce_3^c|\mathbf{x})\le (2\alpha+1)2n^{-1/3}\le n^{-1/6}.
        \end{equation*}
    \end{proof}
    \renewcommand*{\proofname}{Proof of Lemma~\ref{lem:ce4}}  
    \begin{proof}
        Let $\mathbf{x}$ be an input instance with $w(\mathbf{x})=k-1$. For each $j\in\ct$, we have $H_j|\mathbf{x}\sim\Binom(k-1,P_j^1)$. From Lemma~\ref{lem:P1-0}, we know that $P_j^1\le k^{-1-\frac{\gamma}{\log\log(1/\delta)}}$, where $\gamma$ is a positive constant. Therefore, we have $\E[H_j]\le k^{-\frac{\gamma}{\log\log(1/\delta)}}$ for each $j\in\ct$. By Markov's inequality, we know that $\P(H_j\ge 1)\le k^{-\frac{\gamma}{\log\log(1/\delta)}}$ for any $j\in\ct$. Applying a union bound over all $j\in\ct$ yields $$\P(\ce_4|\mathbf{x})\le O\left(\frac{\alpha}{k^\gamma}\right)=O\left(\frac{\log k\log\log(1/\delta)}{k^\frac{\gamma}{\log\log(1/\delta)}}\right).$$
    \end{proof}

    \subsection{Regime 2: \texorpdfstring{$\log k >\log(1/\delta)\log\log(1/\delta)$}{Lg}}
\label{appd:proof-conv-o}
    In this section, we prove Theorem~\ref{thm:thk-conv} in regime
    $\log k >\log(1/\delta)\log\log(1/\delta)$. The proof in this regime is similar to the proof in the previous regime. 
    The major difference is that $\frac{1}{\delta}$ is relatively small comparing to $k$ in this regime, so the statistics exhibit slightly different concentration behaviors.

    Let $\eps=(\log\frac 1\delta)^{-1/4}$ in this regime. Because $\delta=o(1)$ and $\log k=\omega(\log (1/\delta))$,
    we know that $\frac{(1-\epsilon)(n-k-\eps n)\log k}{\div(p-\epsilon\|1-p)}=(1-o(1))\frac{n\log(k/\delta)}{\DKL}$.
    Therefore, it suffices to show that for any algorithm that makes at most $\frac{(1-\epsilon)(n-k-\eps n)\log k}{\div(p-\epsilon\|1-p)}$ queries in expectation, the worst-case error probability is at least $\delta$.
    
    Let $\mathcal{A}$ be an arbitrary algorithm that uses at most $\frac{(1-\epsilon)(n-k-\eps n)\log k}{\div(p-\epsilon\|1-p)}$ queries in expectation. We slightly modify the the procedure of the enhanced algorithm $\mathcal{A}'$ defined in Section~\ref{sec:proof-lower} as follows:
    \begin{itemize}
    \item Non-adaptive phase: query each of the $n$ bits $\alpha\triangleq \frac{(1-\epsilon)\log k}{\div(p-\epsilon\|1-p)}$ times.
    \item Adaptive phase: Adaptively choose $N'$ bits and \emph{noiselessly} reveal their value, where the number of revealed bits $N'$ satisfies $\E[N'|\mathbf{x}]=n-k-\eps n$ for any $\mathbf{x}$.
\end{itemize}

    Following an analogous argument as in Lemma~\ref{lem:enhancement}, we know that the worst-case error probability of $\ca'$ is upper bounded by that of $\ca$. Therefore, it suffices to show that the worst-case error probability of $\ca'$ is at least $\delta$.
    In the analysis of $\mathcal{A}'$, we assume that the input instance satisfies either $w(\mathbf{x})=k-1$ or $w(\mathbf{x})=k$, and there exists genie-aided information in the form of two sequences of sets $\bar{\mathcal{H}}_{0}^ {\alpha}=(\bar{\mathcal{H}}_{0},\ldots, \bar{\mathcal{H}}_{\alpha})$ and $\bar{\mathcal{L}}_{0}^ {\alpha}=(\bar{\mathcal{L}}_{0},\ldots, \bar{\mathcal{L}}_{\alpha})$ defined in the same way as the previous regime. Recall that $\widehat{\sfth}_k({\bf H}_0^\alpha,{\bf L}_0^\alpha,\mathbbm{1}_{\{\ce_1\}})$ denote the output estimator of $\mathcal{A}'$, where $\ce_1$ denote the event that the adaptive phase does not find any heavy ball. Set $\mathcal{N}$ is defined as the collections of all $(2\alpha+2)-$tuples $({\bf h}_0^\alpha,{\bf l}_0^\alpha)$ such that $\widehat{\sfth}_k({\bf h}_0^\alpha,{\bf l}_0^\alpha,1)=0$. 
    
    To prove that $\mathcal{A}'$ has worst-case error probability at least $\delta$, we prove  
    $$\P(\widehat{\sfth}_k=0|w(\mathbf{x})=k)> \delta$$
    under the assumption that
    $$\P(\widehat{\sfth}_k=0|w(\mathbf{x})=k-1)\ge 1-\delta.$$

    In the following, we construct the family $\mathcal{D}$ of realizations such that the typicality and the bounded ratio property introduced in Section~\ref{sec:proof-lower} are satisfied.
    Let $\mathcal{D}$ denote the collection of $(2\alpha+2)-$tuples of natural numbers $({\bf h}_0^\alpha,{\bf l}_0^\alpha)$ such that
    \begin{enumerate}
        \item $\sum_{i=0}^\alpha h_i=k-1$, $\sum_{i=0}^\alpha l_i=n-k+1$;
        \item $\sum_{j\in\ct}l_j\ge (n-k+1)\left(1-2\exp\left(-\sqrt{\log \frac{1}{\delta}}\right)\right)\left(1-\sqrt{\frac{\log n}{n}}\right)$;
        \item $(1-\Delta_j^0)(n-k+1)P_j^0\le l_i\le (1+\Delta_j^0)(n-k+1)P_j^0$ for any $j\in\ct$, where $\Delta_j^0\triangleq\sqrt{\frac{\log n}{(n-k+1)P_j^0}}$;
        \item $(1-\Delta_j^1)(k-1)P_j^1\le h_i\le (1+\Delta_j^1)(k-1)P_j^1$ for any $j\in\ct$, where $\Delta_j^1\triangleq\sqrt{\frac{\log n}{(k-1)P_j^1}}$.
    \end{enumerate}
    Notice that the major difference between the definition of $\mathcal{D}$ and the definition of $\mathcal{C}$ in the previous regime is on the last condition. The reason for this distinction is that in this regime, we have $\E[H_j]=\omega(1)$ instead of $\E[H_j]=o(1)$ for each $j\in \ct$. Therefore, the typical realizations of $H_\lb^{\alpha(p-\eps)}$ would center around its mean rather than having zero values. In the following two  propositions, we state the typicality and the bounded ratio property of $\mathcal{D}$.

        \begin{proposition}
   \label{prop:prob-D} 
   Set $\mathcal{D}$ satisfies that
   \begin{equation}
   \label{eq:prob-D}
       \sum_{({\bf h}_0^\alpha,{\bf l}_0^\alpha)\in\mathcal{D}}\P(\{({\bf H}_0^\alpha,{\bf L}_0^\alpha)=({\bf h}_0^\alpha,{\bf l}_0^\alpha)\}|w(\mathbf{x})=k-1)\ge 1-k^{-1/6}-O(n^{-1/6}).
   \end{equation}
\end{proposition}

\begin{proposition}
\label{prop:pmf-D}
   Let $$n^*\triangleq(n-k+1)\left(1-2\exp\left(-\sqrt{\log \frac{1}{\delta}}\right)\right)\left(1-\sqrt{\frac{\log n}{n}}\right)-(2\eps\alpha+1).$$
    For any $({\bf h}_0^\alpha,{\bf l}_0^\alpha)\in\mathcal{D}$, we have 
    \begin{align}
        &\frac{\P\left(\{(\bar{\bf H}_0^\alpha,\bar{\bf L}_0^\alpha)=({\bf h}_0^\alpha,{\bf l}_0^\alpha)\}\cap\ce_1|w(\mathbf{x})=k\right)}{\P\left(\{(\bar{\bf H}_0^\alpha,\bar{\bf L}_0^\alpha)=({\bf h}_0^\alpha,{\bf l}_0^\alpha)\}|w(\mathbf{x})=k-1\right)}\label{eq:ratio-D}\\
        &\ge\frac{1+o(1)}{n}\left(n^*-\E[N'|w(\bx)=k-1,(\bar{\bf H}_0^\alpha,\bar{\bf L}_0^\alpha)=({\bf h}_0^\alpha,{\bf l}_0^\alpha)]\right).\label{eq:pmf-D}
    \end{align}
\end{proposition}

We are now ready to prove $\P(\widehat{\sfth}_k=0|w(\mathbf{x})=k)> \delta$ based on Propositions~\ref{prop:prob-D} and \ref{prop:pmf-D}.
We have 
\begin{align}
    &\P(\widehat{\sfth}_k=0|w(\mathbf{x})=k)\nonumber\\
    &=\sum_{({\bf h}_0^\alpha,{\bf l}_0^\alpha)\in \mathcal{N}}\P(\{(\bar{\bf H}_0^\alpha,\bar{\bf L}_0^\alpha)=({\bf h}_0^\alpha,{\bf l}_0^\alpha)\}\cap\ce_1|w(\mathbf{x})=k)\nonumber\\
    &\ge \sum_{({\bf h}_0^\alpha,{\bf l}_0^\alpha)\in \mathcal{N}\cap\mathcal{D}}\P(\{(\bar{\bf H}_0^\alpha,\bar{\bf L}_0^\alpha)=({\bf h}_0^\alpha,{\bf l}_0^\alpha)\}\cap\ce_1|w(\mathbf{x})=k)\nonumber\\
    &\stackrel{(a)}{\ge}\sum_{({\bf h}_0^\alpha,{\bf l}_0^\alpha)\in \mathcal{N}\cap\mathcal{D}}\P(\{(\bar{\bf H}_0^\alpha,\bar{\bf L}_0^\alpha)=({\bf h}_0^\alpha,{\bf l}_0^\alpha)\}|w(\mathbf{x})=k-1)\nonumber\\
    &\hspace{1em}\cdot\frac{1+o(1)}{n}\left(n^*-\E[N'|w(\bx)=k-1,(\bar{\bf H}_0^\alpha,\bar{\bf L}_0^\alpha)=({\bf h}_0^\alpha,{\bf l}_0^\alpha)]\right)\nonumber\\
    &=\frac{1+o(1)}{n}\Bigg(\sum_{({\bf h}_0^\alpha,{\bf l}_0^\alpha)\in \mathcal{N}\cap\mathcal{D}}n^*\P(\{(\bar{\bf H}_0^\alpha,\bar{\bf L}_0^\alpha)=({\bf h}_0^\alpha,{\bf l}_0^\alpha)\}|w(\mathbf{x})=k-1)\nonumber\\
    &\hspace{1em} -\P(\{(\bar{\bf H}_0^\alpha,\bar{\bf L}_0^\alpha)=({\bf h}_0^\alpha,{\bf l}_0^\alpha)\}|w(\mathbf{x})=k-1)\E[N'|w(\bx)=k-1,(\bar{\bf H}_0^\alpha,\bar{\bf L}_0^\alpha)=({\bf h}_0^\alpha,{\bf l}_0^\alpha)]\Bigg),\label{eq:two_sums_D}
\end{align}
where (a) follows by Proposition~\ref{prop:pmf-D}.
By Proposition~\ref{prop:prob-D} and the definition of $n^*$, we have
\begin{align}
    &\sum_{({\bf h}_0^\alpha,{\bf l}_0^\alpha)\in \mathcal{N}\cap\mathcal{D}}n^*\P(\{(\bar{\bf H}_0^\alpha,\bar{\bf L}_0^\alpha)=({\bf h}_0^\alpha,{\bf l}_0^\alpha)\}|w(\mathbf{x})=k-1)\nonumber\\
    &\ge n^*(1-\delta-k^{-1/6}-O(n^{-1/6}))\nonumber\\
    &=(n-k)\left(1-\delta-O(k^{-1/6})-O\left(\exp\left(-\sqrt{\log\tfrac{1}{\delta}}\right)\right)-O\left(\sqrt{\tfrac{\log n}{n}}\right)-O\left(\frac{\eps\alpha}{n}\right)\right).\label{eq:sum1-D}
\end{align}
By the law of total expectation and the fact that $N'$ is always non-negative, we have
\begin{align}
    &\sum_{({\bf h}_0^\alpha,{\bf l}_0^\alpha)\in \mathcal{N}\cap\mathcal{D}}\P(\{(\bar{\bf H}_0^\alpha,\bar{\bf L}_0^\alpha)=({\bf h}_0^\alpha,{\bf l}_0^\alpha)\}|w(\mathbf{x})=k-1)\E[N'|w(\bx)=k-1,(\bar{\bf H}_0^\alpha,\bar{\bf L}_0^\alpha)=({\bf h}_0^\alpha,{\bf l}_0^\alpha)]\nonumber\\
    &\le\E[N'|w(\bx)=k-1]\le n-k-\eps n.\label{eq:sum2-D}
\end{align}
\begin{sloppypar}
Now, we focus on the difference between \eqref{eq:sum1-D} and \eqref{eq:sum2-D}. Recall that we defined $\eps=(\log\frac 1\delta)^{-1/4}$. It follows that
\begin{align*}
    &(n-k)\left(1-\delta-O(k^{-1/6})-O\left(\exp\left(-\sqrt{\log\tfrac{1}{\delta}}\right)\right)-O\left(\sqrt{\tfrac{\log n}{n}}\right)-O\left(\frac{\eps\alpha}{n}\right)\right)-(n-k-\eps n)\\
    &\ge(1-o(1))\epsilon n.
\end{align*}
\end{sloppypar}
Therefore,~\eqref{eq:two_sums_D} implies that
\begin{equation*}
    \P(\widehat{\sfth}_k=0|w(\mathbf{x})=k)\ge (1-o(1))\epsilon.
\end{equation*}
Notice that $\frac{\eps}{\delta}\ge\frac{(\log\tfrac{1}{\delta})^{-1/4}}{\delta}=\omega(1)$. This shows that 
\begin{equation*}
    \P(\widehat{\sfth}_k=0|w(\mathbf{x})=k)\ge\omega(\delta).
\end{equation*}

Now it suffices to prove Propositions~\ref{prop:prob-D} and \ref{prop:pmf-D} to complete the proof in this regime.

\subsubsection{Proof of Propositions}
\renewcommand*{\proofname}{Proof of Proposition \ref{prop:prob-D}}  
\begin{proof}
    Define error events
    $$\ce_5\triangleq\left\{\sum_{j\in\ct}\le (n-k+1)\left(1-2\exp\left(-\sqrt{\log \frac{1}{\delta}}\right)\right)\left(1-\sqrt{\frac{\log n}{n}}\right)\right\},$$
    $$\ce_6\triangleq\{\exists j\in\ct:L_j\notin[(1-\Delta_j^0)(n-k+1)P_j^0,(1+\Delta_j^0)(n-k+1)P_j^0]\},$$
    and
    $$\ce_7\triangleq\{\exists j\in\ct:H_j\notin[(1-\Delta_j^1)(k-1)P_j^1,(1+\Delta_j^1)(k-1)P_j^1]\}.$$
    Notice that events $\ce_5,\ce_6$ and $\ce_7$ correspond to the second, third and fourth conditions in the definition of $\mathcal{D}$. It follows that
    \begin{align*}
        &\sum_{({\bf h}_0^\alpha,{\bf l}_0^\alpha)\in\mathcal{D}}\P(\{({\bf H}_0^\alpha,{\bf L}_0^\alpha)=({\bf h}_0^\alpha,{\bf l}_0^\alpha)\}|w(\mathbf{x})=k-1)\\
            &= 1-\P(\ce_5\cup\ce_6\cup\ce_7|w(\mathbf{x})=k-1)\\
            &\ge 1-\P(\ce_5|w(\mathbf{x})=k-1)-\P(\ce_6|w(\mathbf{x})=k-1)-\P(\ce_7|w(\mathbf{x})=k-1),
    \end{align*}
    where the last inequality follows by the union bound. Proposition~\ref{prop:prob-D} follows readily by the following three lemmas.
            \begin{lemma}
        \label{lem:ce5}
        For any $\mathbf{x}$ with $w(\mathbf{x})=k-1$, we have
        $\P(\ce_5|\mathbf{x})\le n^{-1/5}.$
        \end{lemma}
        \begin{lemma}
    \label{lem:ce6}
        For any $\mathbf{x}$ with $ w(\mathbf{x})=k-1$, we have $\P(\ce_6|\mathbf{x})\le n^{-1/6}$.
    \end{lemma}
    \begin{lemma}
    \label{lem:ce7}
        For any $\mathbf{x}$ with $ w(\mathbf{x})=k-1$, we have $\P(\ce_7|\mathbf{x})\le k^{-1/6}$.
    \end{lemma}
    By Lemmas~\ref{lem:ce5}, \ref{lem:ce6} and \ref{lem:ce7}, we have
    \begin{align*}
        &\sum_{({\bf h}_0^\alpha,{\bf l}_0^\alpha)\in\mathcal{D}}\P(\{({\bf H}_0^\alpha,{\bf L}_0^\alpha)=({\bf h}_0^\alpha,{\bf l}_0^\alpha)\}|w(\mathbf{x})=k-1)\\
        &\ge 1-O(n^{-1/6})-k^{-1/6},
    \end{align*}
    which completes the proof.
\end{proof}

\renewcommand*{\proofname}{Proof of Proposition~\ref{prop:pmf-D}}  
\begin{proof}
    Following the same derivation for equation~\eqref{eq:pmf-ratio} in the proof of Proposition~\ref{prop:pmf-C}, we have
    \begin{align}
        &\frac{\P\left(\{(\bar{\bf H}_0^\alpha,\bar{\bf L}_0^\alpha)=({\bf h}_0^\alpha,{\bf l}_0^\alpha)\}\cap\ce_1|w(\mathbf{x})=k\right)}{\P\left(\{(\bar{\bf H}_0^\alpha,\bar{\bf L}_0^\alpha)=({\bf h}_0^\alpha,{\bf l}_0^\alpha)\}|w(\mathbf{x})=k-1\right)}\nonumber\\
        &=\sum_{j\in\ct}\frac{kP_j^1}{h_j+1}\cdot\frac{l_j}{(n-k+1)P_j^0}\cdot\frac{l_j-1-\E[N'_j|\bar{\bf H}_0^\alpha={\bf h}_0^\alpha,\bar{\bf L}_0^\alpha={\bf l}_0^\alpha,w(\mathbf{x})=k-1]}{-1+\sum_{i\in\ct} l_i}.\label{eq:pmf-ratio-D}
    \end{align}
    To proceed, we need a technical lemma that bounds $P_j^0$ and $P_j^1$ for each $j\in\ct$.
    \begin{lemma}
        \label{lem:P1-0-o}
        For each $j\in\ct$, we have $P_j^1\ge\exp\left(-\log k+\frac{\log k}{2(\log\tfrac1\delta)^{1/4}}\right)$ and $P_j^0\ge\exp(-o(\log k))$.
    \end{lemma}
    Because $({\bf h}_0^\alpha,{\bf l}_0^\alpha)\in\mathcal{C}$, we have $(1-\Delta_j^1)(k-1)P_j^1\le h_i\le (1+\Delta_j^1)(k-1)P_j^1$ and $(1-\Delta_j^0)(n-k+1)P_j^0\le l_i\le (1+\Delta_j^0)(n-k+1)P_j^0$ for each $j\in\ct$. Recall that we defined $\Delta_j^1\triangleq\sqrt{\frac{\log k}{(k-1)P_j^1}}$ and $\Delta_j^0\triangleq\sqrt{\frac{\log n}{(n-k+1)P_j^0}}$. By Lemma~\ref{lem:P1-0-o}, we have $\Delta_j^0=o(1)$ and $\Delta_j^1=o(1)$, and it follows that 
    \begin{equation*}
        \frac{kP_j^1}{h_j+1}=1+o(1)\;\text{ and }\;\frac{l_j}{(n-k+1)P_j^0}=1+o(1).
    \end{equation*}
    Therefore, we have
    \begin{align*}
        &\frac{\P\left(\{(\bar{\bf H}_0^\alpha,\bar{\bf L}_0^\alpha)=({\bf h}_0^\alpha,{\bf l}_0^\alpha)\}\cap\ce_1|w(\mathbf{x})=k\right)}{\P\left(\{(\bar{\bf H}_0^\alpha,\bar{\bf L}_0^\alpha)=({\bf h}_0^\alpha,{\bf l}_0^\alpha)\}|w(\mathbf{x})=k-1\right)}\\
        &=(1+o(1))\sum_{j\in\ct}\frac{l_j-1-\E[N'_j|\bar{\bf H}_0^\alpha={\bf h}_0^\alpha,\bar{\bf L}_0^\alpha={\bf l}_0^\alpha,w(\mathbf{x})=k-1]}{-1+\sum_{i\in\ct} l_i}\\
        &\ge(1+o(1))\frac{-\E[N'|\bar{\bf H}_0^\alpha={\bf h}_0^\alpha,\bar{\bf L}_0^\alpha={\bf l}_0^\alpha,w(\mathbf{x})=k-1]-(2\eps\alpha+1)+\sum_{j\in\ct}l_j}{-1+\sum_{i\in\ct} l_i}.
    \end{align*}
    Because $({\bf h}_0^\alpha,{\bf l}_0^\alpha)\in\mathcal{D}$, we have 
    \[
    \sum_{i\in\ct} l_i\ge (n-k+1)\left(1-2\exp\left(-\sqrt{\log \frac{1}{\delta}}\right)\right)\left(1-\sqrt{\frac{\log n}{n}}\right).
    \]
    Finally, by the definition of $n^*$ in the proposition statement, we have
    \begin{align*}
        &\frac{\P\left(\{(\bar{\bf H}_0^\alpha,\bar{\bf L}_0^\alpha)=({\bf h}_0^\alpha,{\bf l}_0^\alpha)\}\cap\ce_1|w(\mathbf{x})=k\right)}{\P\left(\{(\bar{\bf H}_0^\alpha,\bar{\bf L}_0^\alpha)=({\bf h}_0^\alpha,{\bf l}_0^\alpha)\}|w(\mathbf{x})=k-1\right)}\\
        &\ge \frac{1+o(1)}{n}\left(n^*-\E[N'|w(\bx)=k-1,(\bar{\bf H}_0^\alpha,\bar{\bf L}_0^\alpha)=({\bf h}_0^\alpha,{\bf l}_0^\alpha)]\right),
    \end{align*}
    which completes the proof.
\end{proof}

\subsubsection{Proof of Lemmas}
\renewcommand*{\proofname}{Proof of Lemma~\ref{lem:P1-0-o}}  
\begin{proof}
    By definition, we have $P_j^1=\binom{\alpha}{j}(1-p)^jp^{\alpha-j}$.
    Notice that $\min_{j\in\ct}P_j^1=P_\lb^1$. Therefore, it suffices to show the claim for $j=\lb$.
        By the pmf of binomial distribution and Stirling's approximation, we have
        \begin{align*}
            P_\lb^1&=\binom{\alpha}{(p-\eps)\alpha}(1-p)^{\alpha(p-\eps)}p^{\alpha(1-p+\eps)}\\
            &\ge \exp\left(-\alpha\left(\div(p-\eps\|1-p)+O\left(\frac{\log\alpha}{\alpha}\right)\right)\right)\\
            &=\exp\left(-\frac{(1-\eps)\log k}{\div(p-\eps\|1-p)}\left(\div(p-\eps\|1-p)+O\left(\frac{\log\alpha}{\alpha}\right)\right)\right)\\
            &\stackrel{(a)}{\ge} \exp\left(-\left(1-\frac{1}{(\log\tfrac 1\delta)^{1/4}}\right)\left(1+O\left(\frac{\log\log k}{\log k}\right)\right)\log k\right)\\
            &\stackrel{(b)}{\ge} \exp\left(-\log k+\frac{\log k}{2(\log\tfrac1\delta)^{1/4}}\right),
        \end{align*}
        where (a) follows because $\eps= (\log\tfrac{1}{\delta})^{-1/4}$ and $\alpha=\Theta(\log k)$, and (b) follows because $\frac{1}{(\log\tfrac 1\delta)^{1/4}}=\omega\left(\frac{\log\log k}{\log k}\right)$.

        We have $P_j^0=\binom{\alpha}{j}p^j(1-p)^{\alpha-j}$. We can see that $\min_{j\in\ct}P_j^0=\min(P_{\alpha(p-\eps)}^0,P_{\alpha(p+\eps)}^0)$. Therefore, it suffices to show that $P_{\alpha(p-\eps)}^0\ge \exp(-o(\log k))$ and $P_{\alpha(p+\eps)}^0\ge \exp(-o(\log k))$. 
        By Stirling's approximation, we have
        \begin{align*}
            P_{\alpha(p-\eps)}^0&=\binom{\alpha}{\alpha(p-\eps)}p^{\alpha(p-\eps)}(1-p)^{\alpha(1-p+\eps)}\\
            &\ge \exp\left(-\alpha\left(\div(p-\eps\|p)+O\left(\frac{\log\alpha}{\alpha}\right)\right)\right)\\
            &=\exp\left(-\frac{(1-\eps)\log k}{\div(p-\eps\|1-p)}\left(\div(p-\eps\|p)+O\left(\frac{\log\alpha}{\alpha}\right)\right)\right)\\
            &\ge \exp(-o(\log k)),
        \end{align*}
        where the last inequality follows because $\div(p-\eps\|p)=o(1)$.
        It can be similarly shown that $P_{\alpha(p+\eps)}^0\ge \exp(-o(\log k))$.
\end{proof}
\renewcommand*{\proofname}{Proof of Lemma~\ref{lem:ce5}}
\begin{proof}
    For any $\mathbf{x}$ with $w(\mathbf{x})=k-1$, we know that $\sum_{j\in\ct}L_j|\mathbf{x}\sim \Binom(n-k+1,\P(W_0\in\ct))$, where $W_0\sim\Binom(\alpha,p)$. To bound $\P(\ce_5|\mathbf{x})$, we first bound $\P(W_0\in\ct)$. We have
        \begin{align*}
    \P(W_0\notin\ct)&=\P(|W_0-\alpha p|\ge \alpha\eps)\\
    &\stackrel{(a)}{\le}2\exp\left(-\frac{\eps^2\alpha}{3p}\right)\\
    &\stackrel{(b)}{\le}2\exp\left(-\frac{(1-\eps)\log k}{3p\div(p-\eps\|1-p)\sqrt{\log(1/\delta)}}\right)\\
    &\stackrel{(c)}{\le} 2\exp\left(-\sqrt{\log \frac{1}{\delta}}\right),
\end{align*}
    where (a) follows by the Chernoff bound, (b) follows because $\eps= (\log (1/\delta))^{-1/4}$, and (c) follows because $\log k=\omega(\log\frac 1\delta)$.

        Therefore, we have
    \begin{align*}
        &\P(\ce_5|\mathbf{x})\\
        &\le \P\Bigg(\mathrm{Binom}\left(n-k+1,1-2\exp\left(-\sqrt{\log \tfrac{1}{\delta}}\right)\right)\\
        &\hspace{2em}\le (n-k+1)\left(1-2\exp\left(-\sqrt{\log \tfrac{1}{\delta}}\right)\right)\left(1-\sqrt{\tfrac{\log n}{n}}\right)\Bigg)\\
        &\stackrel{(d)}{\le} \exp\left(-\frac{(n-k+1)\left(1-2\exp\left(-\sqrt{\log \frac{1}{\delta}}\right)\right)\log n}{2n}\right)\\
        &\le n^{-1/5},
    \end{align*}
    where (d) follows by the Chernoff bound.
\end{proof}
\renewcommand*{\proofname}{Proof of Lemma~\ref{lem:ce6}}
\begin{proof}
    For any $\mathbf{x}$ with $w(\mathbf{x})=k-1$, we have $L_j|\mathbf{x}\sim\Binom(n-k+1,P_j^0)$ for each $j\in\ct$. For any $j\in\ct$, by Chernoff bound we have
        \begin{align*}
            &\P(|L_j-(n-k+1)P_j^0|\ge \Delta_j^0(n-k+1)P_j^0|\mathbf{x})\\
            &\le 2\exp(-(\Delta_j^0)^2(n-k+1)P_j^0/3)\\
            &=2n^{-1/3},
        \end{align*}
    where the last equality follows because $\Delta_j^0=\sqrt{\frac{\log n}{(n-k+1)P_j^0}}$. By union bound, we have
        \begin{equation*}
            \P(\ce_6^c|\mathbf{x})\le (2\alpha+1)2n^{-1/3}\le n^{-1/6}.
        \end{equation*}    
\end{proof}
\renewcommand*{\proofname}{Proof of Lemma~\ref{lem:ce7}}
\begin{proof}
    For any $\mathbf{x}$ with $w(\mathbf{x})=k-1$, we have $H_j|\mathbf{x}\sim\Binom(k-1,P_j^1)$ for each $j\in\ct$. For any $j\in\ct$, by Chernoff bound we have
        \begin{align*}
            &\P(|H_j-(k-1)P_j^1|\ge \Delta_j^1((k-1)P_j^1|\mathbf{x})\\
            &\le 2\exp(-(\Delta_j^1)^2(k-1)P_j^1/3)\\
            &=2k^{-1/3},
        \end{align*}
    where the last equality follows because $\Delta_j^1=\sqrt{\frac{\log k}{(k-1)P_j^1}}$. By union bound, we have
        \begin{equation*}
            \P(\ce_7^c|\mathbf{x})\le (2\alpha+1)2k^{-1/3}\le k^{-1/6}.
        \end{equation*} 
\end{proof}

\subsection{Regime 3: \texorpdfstring{$\log k\le \frac{\log(1/\delta)}{\log\log(1/\delta)}$}{Lg}}
\label{appd:proof-conv-omega}
    In this regime, we prove the converse using an approach called Le Cam's two points methods (see, e.g.~\cite{yu1997assouad}). The proof is done by establishing the following proposition.
    \begin{proposition}
    \label{prop:le-cam}
        For any variable-length algorithm for computing $\sfth_k(\mathbf{x})$ with the number of queries $M$ satisfying
        \begin{equation*}
            \E[M|\bx]\le \frac{(n-k+1)\log\tfrac{1}{4\delta}}{\DKL}
        \end{equation*}
        for any input instance $\bx$. Then the worst-case error probability of the algorithm is at least $\delta$.
    \end{proposition}
    Notice that under assumptions $\delta=o(1)$ and $\log k\le \frac{\log(1/\delta)}{\log\log(1/\delta)}$, we have 
    \[
    \frac{(n-k+1)\log\tfrac{1}{4\delta}}{\DKL}=(1-o(1))\frac{(n-k)\log\frac{k}{\delta}}{\DKL}.
    \]
    Hence Proposition~\ref{prop:le-cam} directly implies Theorem~\ref{thm:thk-conv} in this regime. It is worth noting that Proposition~\ref{prop:le-cam} does not imply the same bounds in the other two regimes, namely when $\frac{\log(1/\delta)}{\log\log(1/\delta)}<\log k \le \log(1/\delta)\log\log(1/\delta)$ and when $\log k >\log(1/\delta)\log\log(1/\delta)$.

    The proof of Proposition~\ref{prop:le-cam} is based on Le Cam's Two Points Lemma stated below.
    \begin{lemma}[Le Cam's Two Points Lemma~\citep{yu1997assouad}]\label{lem:lecam}
    For any $\theta_0,\theta_1\in\Theta$, suppose that the following separation condition holds for some loss function $L(\theta, a):\Theta\times \mathcal{A}\rightarrow \mathbb{R}$:
    \begin{align*}
    \forall {a\in\mathcal{A}}, L(\theta_0, a)+L(\theta_1, a)\geq \Delta>0.
    \end{align*}
    Then we have
    \begin{align*}
        \inf_f \sup_\theta \mathbb{E}_\theta[L(\theta, f(X))]\geq \frac{\Delta}{2}\left(1-\mathsf{TV}(\mathbb{P}_{\theta_0},\mathbb{P}_{\theta_1})\right), 
    \end{align*}
    where $\mathbb{P}_{\theta_0}$ and $\mathbb{P}_{\theta_1}$ represent the probability distribution of the statistics $X$ under the two hypotheses $\theta_1$ and $\theta_2$ respectively.
\end{lemma}
\renewcommand*{\proofname}{Proof of Proposition~\ref{prop:le-cam}}
    \begin{proof}
        To apply Lemma~\ref{lem:lecam}, we first define $n-k+2$ input instances for the $\sfth_k$ problem. 
        Let $\mathbf{x}^{(0)}$ denote the binary sequence such that
        \begin{equation*}
            x_i^{(0)}=\begin{cases}
                1, &\text{ for }1\le i\le k-1,\\
                0, &\text{ otherwise}.
            \end{cases}
        \end{equation*}
        For each $k\le j\le n$, let $\mathbf{x}^{(j)}$ denote the binary sequence such that 
        \begin{equation*}
            x_i^{(j)}=\begin{cases}
                1, &\text{ for }1\le i\le k-1 \text{ or }i=j,\\
                0, &\text{ otherwise}.
            \end{cases}
        \end{equation*}
        Notice that $\sfth_k(\mathbf{x}^{(0)})=0$ and $\sfth_k(\mathbf{x}^{(j)})=1$ for each $k\le j\le n$. Therefore, for any estimator $\widehat{\sfth}_k$, we have
        \begin{align*}
    \mathbbm{1}(\widehat{\sfth}_k \neq \mathsf{TH}_k(\mathbf{x}^{(0)})) +   \mathbbm{1}(\widehat{\sfth}_k \neq \mathsf{TH}_k(\mathbf{x}^{(j)})) \geq 1.
\end{align*}
        By Lemma~\ref{lem:lecam}, we have 
        \begin{align*}
      \inf_{\widehat{\sfth}_k} \sup_{\mathbf{x}\in\{0, 1\}^K} \P(\widehat{\sfth}_k \neq \mathsf{TH}_k(\mathbf{x})) & \geq \frac{1}{2}(1-\mathsf{TV}(\mathbb{P}_{\mathbf{x}^{(0)}}, \mathbb{P}_{\mathbf{x}^{(j)}})),
        \end{align*}
        for any $k\le j\le n$, where $\mathbb{P}_{\mathbf{x}^{(0)}}$ denote the joint distribution of query-observation pairs in $m$ rounds when the ground truth is $\mathbf{x}^{(0)}$ and $\mathbb{P}_{\mathbf{x}^{(j)}}$ denote the joint distribution of query-observation pairs in $m$ rounds when the ground truth is $\mathbf{x}^{(j)}$. We can take maximum over $k\le j\le n$ on the right-hand side and get
        \begin{align*}
             &\inf_{\widehat{\sfth}_k} \sup_{\mathbf{x}\in\{0, 1\}^K} \P(\widehat{\sfth}_k \neq \mathsf{TH}_k(\mathbf{x}))\\
             &\ge \sup_{k\le j\le n} \frac{1}{2}(1-\mathsf{TV}(\mathbb{P}_{\mathbf{x}^{(0)}}, \mathbb{P}_{\mathbf{x}^{(j)}}))\\
             &\ge \sup_{k\le j\le n} \frac{1}{4}\exp(-\div(\mathbb{P}_{\mathbf{x}^{(0)}}, \mathbb{P}_{\mathbf{x}^{(j)}})),             
        \end{align*}
        where the last inequality follows by the Bretagnolle–Huber inequality stated in Lemma~\ref{lem:bh}. Now, we need to bound the KL divergence $\div(\mathbb{P}_{\mathbf{x}^{(0)}}, \mathbb{P}_{\mathbf{x}^{(j)}})$. By the Divergence Decomposition Lemma stated in Lemma~\ref{lem:div-decomp}, we have
        \begin{equation*}
            \div(\mathbb{P}_{\mathbf{x}^{(0)}}, \mathbb{P}_{\mathbf{x}^{(j)}})=\E[M_j|\bx^{(0)}]\DKL,
        \end{equation*}
        where $M_j$ denote the number of queries on the bit $x_j$. Because $\sum_{j=k}^n \E[M_j|\bx^{(0)}]\le \E[M|\bx^{(0)}]\le \frac{(n-k+1)\log\tfrac{1}{4\delta}}{\DKL}$, there exists some $k\le j^*\le n$ such that $\E[M_j|\bx^{(0)}]\le \frac{\log\tfrac{1}{4\delta}}{\DKL}$. Finally, we have
        \begin{align*}
            &\inf_{\widehat{\sfth}_k} \sup_{\mathbf{x}\in\{0, 1\}^K} \P(\widehat{\sfth}_k \neq \mathsf{TH}_k(\mathbf{x}))\\
            &\ge \sup_{k\le j\le n} \frac{1}{4}\exp(-\E[M_j|\bx^{(0)}]\DKL)\\
            &\ge \frac{1}{4}\exp(-\E[M_{j^*}|\bx^{(0)}]\DKL)\\
            &\ge\delta,
        \end{align*}
        which completes the proof.
    \end{proof}
\begin{lemma}[An upper Bound of Bretagnolle–Huber inequality
 \citep{bretagnolle79}]\label{lem:bh}
    For any distribution $\mathbb{P}_1,\mathbb{P}_2$, one has
    \begin{align*}
        \mathsf{TV}(\mathbb{P}_1, \mathbb{P}_2)\leq 1-\frac{1}{2} \exp(-D_{\mathsf{KL}}(\mathbb{P}_1, \mathbb{P}_2)).
    \end{align*}
\end{lemma}

\begin{lemma}[Divergence Decomposition \citep{auer1995gambling}]\label{lem:div-decomp}
Let $T_i$ be the  random variable denoting the number of times experiment $i\in[K]$ is performed under some policy $\pi$, then for two distributions $\mathbb{P}^\pi, \mathbb{Q}^\pi$ under policy $\pi$,
\begin{align*}
    D_{\mathsf{KL}}(\mathbb{P}^\pi, \mathbb{Q}^\pi) = \sum_{i\in[K]}\mathbb{E}_{\mathbb{P}^\pi}[T_i] D_{\mathsf{KL}}(\mathbb{P}_i^\pi, \mathbb{Q}_i^\pi).
\end{align*}
\end{lemma}

\section{Detailed Proof of Theorem~\ref{thm:threshold_achievability}} \label{appd:proof-ach}
In this appendix, we give a complete proof of Theorem~\ref{thm:threshold_achievability}. We present the proof in two separate regimes $k\le\frac{n}{\log n}$ and $k>\frac{n}{\log n}$.
\subsection{Regime 1: \texorpdfstring{$k\le \frac{n}{\log n}$}{Lg}}
To prove Theorem~\ref{thm:threshold_achievability} in this regime, we establish the following proposition.
\begin{proposition}
\label{prop:upper_regime1}
    Suppose $k\le \frac{n}{\log n}$ and $k\le n/2$. There exists a variable-length algorithm, namely Algorithm~\ref{alg:proposed}, that computes the $\sfth_k$ function with worst-case error probability at most $3\delta$. Moreover, the number of queries $M$ made by the algorithm satisfies 
    $$\E[M|\bx]\le\frac{n\log\frac{k}{\delta}}{\DKL}+\frac{n}{1-2p}+C\left(k+\max\left(n\delta+n\sqrt{\delta},\frac{n}{\log n}\right)\right)\log\frac{k}{\delta}$$
    for any input instance $\bx$.
\end{proposition}
To prove Theorem~\ref{thm:threshold_achievability} in this regime, we can set $\delta$ in Proposition~\ref{prop:upper_regime1} to $\delta'/3$. Then we know that Algorithm~\ref{alg:proposed} has worst-case error probability at most $\delta'$, and its expected query complexity satisfies
\[
\E[M|\bx]\le\frac{n\log\frac{3k}{\delta'}}{\DKL}+\frac{n}{1-2p}+C\left(k+\max\left(n\delta'/3+n\sqrt{\delta'/3},\frac{n}{\log n}\right)\right)\log\frac{3k}{\delta'}.
\]
By the assumptions that $k\le \frac{n}{\log n}$ and $\delta'=o(1)$, it follows that
\[
\E[M|\bx]=(1+o(1))\frac{n\log\frac{k}{\delta'}}{\DKL},
\]
which proves Theorem~\ref{thm:threshold_achievability} in this regime.

In the rest of this section, we prove Proposition~\ref{prop:upper_regime1} by analyzing the proposed \textsc{NoisyThreshold} algorithm (Algorithm~\ref{alg:proposed}).

Recall that Algorithm~\ref{alg:proposed} uses the \textsc{CheckBit} and \textsc{MaxHeapThreshold} functions as subroutines. The \textsc{CheckBit} algorithm takes as input a single bit and returns an estimate of the bit through repeated queries. The \textsc{CheckBit} algorithm is given in Algorithm~\ref{alg:checkbit}, where the parameter $\alpha$ in the $i$th iteration corresponds to the probability that the input bit is one given the first $i$ noisy observations. The following lemma gives an upper bound on the expected number of queries used by the \textsc{CheckBit} algorithm, which can be seen as a recast of Lemma 13 in~\citet{gu2023optimal}.
\setcounter{AlgoLine}{0}
\begin{algorithm}[t]
    \LinesNumbered
    \DontPrintSemicolon
    \caption{\textsc{CheckBit} algorithm for finding the value of an input bit}
    \label{alg:checkbit}
    \KwData{Input bit $x$, tolerated error probability $\delta$, crossover probability $p$.}
    \KwResult{Estimate of $x$.}
    $\alpha\gets 1/2$\;
    
    \While{$\alpha\in(\delta, 1-\delta)$}{
        Query the input bit $x$\;
        
        \eIf{$1$ \text{is observed}}{
            $\alpha \gets \frac{(1-p)\alpha}{(1-p)\alpha + p(1-\alpha)}$\;
        }{
            $\alpha \gets \frac{p\alpha}{p\alpha + (1-p)(1-\alpha)}$\;
        }
    }
    \eIf{$\alpha\ge 1-\delta$}{
        \Return $1$\;
    }{
        \Return $0$\;
    }
\end{algorithm}

\begin{lemma}[Lemma 13 in~\cite{gu2023optimal}] \label{lemma:check-bit}
    There exists a randomized algorithm, namely, the \textsc{CheckBit} algorithm, that finds the value of any input bit $x$ with error probability at most $\delta$. The algorithm makes at most $\frac{1}{1-2p}\left\lceil\frac{\log\frac{1-\delta}{\delta}}{\log\frac{1-p}{p}}\right\rceil$
    queries in expectation.
\end{lemma}

The \textsc{MaxHeapThreshold} algorithm computes the $\sfth_k$ function for a length-$n$ input bit sequence with a desired error probability $\delta$. This algorithm has been previously proposed in~\cite{feige1994computing}, and is given here for completion in Algorithm~\ref{alg:heap_threshold}. The \textsc{MaxHeapThreshold} algorithm uses the following three subroutines.
\begin{itemize}
    \item \textsc{NoisyCompareUsingReadings}~(Algorithm~\ref{alg:noisycomparereading}): which compares two bits using $C_1(p)$ noisy readings of each bit.
    
    \item \constructheap~(Algorithm~\ref{alg:constructheap}): which returns a max-heap representation ${\bf z}$ of ${\bf x}$ such that $z_{i,j}$ is the $j$-th entry in the $i$-th level of a max-heap. The algorithm is similar to the tournament algorithm for computing the \textsc{Max} function~\cite[Algorithm 5]{zhu2023ormax}.
    
    \item \textsc{NoisyExtractMax}: which is similar to the common \textsc{ExtractMax}() operation in a max-heap structure, except that every comparison is performed using $C_2(p)$ noisy readings. Since this is a simple modification to a standard max-heap operation, we don't show this algorithm here.
\end{itemize}

\begin{remark}
    Note that $C_1(p)$ and $C_2(p)$ are constants that depend only on $p$. For the sake of the analysis of our proposed algorithm, the exact value of these constants does not matter.
\end{remark}

The following lemma gives an upper bound on the expected number of queries made by the \textsc{MaxHeapThreshold} algorithm.

\begin{lemma}[Theorem 3.3 in~\cite{feige1994computing}]
\label{lemma:existing_threshold}
    For any positive integer $k<n$ and positive real number $\delta<1$, the \textsc{MaxHeapThreshold} algorithm computes the $\sfth_k$ function with error probability at most $\delta$ and using at most $Cn\log\frac{m}{\delta}$ queries, where $m=\min\{k,n-k+1\}$ and $C$ is a constant depending only on $p$.
\end{lemma}
\setcounter{AlgoLine}{0}
\begin{algorithm}[t]
    \LinesNumbered
    \DontPrintSemicolon
    \caption{\textsc{NoisyCompareUsingReadings} for comparing two elements using noisy readings} \label{alg:noisycomparereading}
    \KwData{Elements $x_1$ and $x_2$, tolerated error probability $\delta$, crossover probability $p$.}
    \KwResult{Estimate of $\max\{x_1,x_2\}$.}
    Make $C_1(p)\log\tfrac1\delta$ noisy readings of $x_1$. Let $\hat{x}_1$ be estimate pertaining to majority of readings.\;
    
    Make $C_1(p)\log\tfrac1\delta$ noisy readings of $x_2$. Let $\hat{x}_2$ be estimate pertaining to majority of readings.\;
    
    \Return $\max\{\hat{x}_1,\hat{x}_2\}$\;
\end{algorithm}
\setcounter{AlgoLine}{0}
\begin{algorithm}[t]
    \LinesNumbered
    \DontPrintSemicolon
    \caption{\constructheap algorithm}
    \label{alg:constructheap}
    \KwData{Sequence ${\bf x} = (x_1,\ldots,x_n)$, tolerated error probability $\delta$, crossover probability $p$.}
    \KwResult{Returns a max-heap structure of ${\bf x}$.}
    Set ${\bf y} \gets {\bf x}$, ${\bf z}_{0,1:n} \gets {\bf y}$, $r \gets n$\;
    
    \For{$i= 1:\lceil\log_2(n)\rceil $}{
        $\tilde\delta_i \gets \delta^{2(2i-1)}$\;
        
        \For{$j = 1:\lfloor r/2\rfloor$}{
            $z_{i,j} \gets \textsc{NoisyCompareUsingReadings}$($y_{2j-1}$, $y_{2j}$, $\tilde\delta_i$, $p$)\;
        }
        \eIf{$r$ is even}{
            ${\bf y} \gets (z_{i,1}, \ldots, z_{i,r/2})$\;
        }{
            ${\bf y} \gets (z_{i,1}, \ldots, z_{i,\lfloor r/2\rfloor}, y_r)$\;
            
            $z_{i,\lceil r/2\rceil} \gets y_r$\;
        }
        $r\gets \lceil r/2 \rceil$\;
    }
    \Return ${\bf z}$\;
\end{algorithm}
\setcounter{AlgoLine}{0}
\begin{algorithm}[t]
    \LinesNumbered
    \DontPrintSemicolon
    \caption{\textsc{MaxHeapThreshold} algorithm for computing the $\sfth_k$ function}
    \label{alg:heap_threshold}
    \KwData{Bit sequence ${\bf x}\hspace*{-0.2em}=\hspace*{-0.2em}(x_1,\hspace*{-0.1em}\ldots\hspace*{-0.1em},x_n)$, parameter $k\leq \frac{n}{2}$, error probability $\delta$, noise probability $p$.}
    \KwResult{Estimate of $\sfth_k({\bf x})$.}
    $y \gets x$\;
    
    \eIf{$k\geq \sqrt{n}$}{
        Apply the sorting algorithm from~\citet{feige1994computing} and return the $k$th largest element\footnotemark{}\;
    }{
        ${\bf y} \gets \constructheap({\bf x}, \,p, \, \frac{\delta}{2k})$\;
        
        \For{$j=1,\ldots,k$}{
            $z_j \gets \textsc{NoisyExtractMax}({\bf y},\, p,\, \frac{\delta}{2k\log n})$\;
        }
        \Return $z_k$\;
    }
\end{algorithm}

By symmetry, we only need to consider the case of $k\leq n/2$. In the following, we analyze the error probability and the number of queries of Algorithm~\ref{alg:proposed}.

\vspace{0.5em}
\emph{\textbf{Error Analysis:}}
The following lemma bounds the worst-case error probability of Algorithm~\ref{alg:proposed}.

\begin{lemma}[Error probability of \textsc{NoisyThreshold}] \label{lemma:error-prob-threshold}
    The worst-case error probability of the \textsc{NoisyThreshold} algorithm is at most $3\delta$.
\end{lemma}

\renewcommand*{\proofname}{Proof}
\begin{proof}
    First, suppose that $\sfth_k({\bf x}) = 1$. Without loss of generality, assume that $x_1 = \cdots = x_k = 1$. Let $\ce_1=\{\widehat{\sfth}_k = 0\}$ denote the error event in this case, and let $\ca = \{\exists \, i\in[k]: x_i \notin \cs\}$. By the union bound, we have
    \[
    \P(\ce_1) \leq \P(\ca) + \P(\ce_1 \cond \ca^c).
    \]
    For the first term, by Lemma~\ref{lemma:check-bit}, we have that $\P(\ca) \leq \sum_{i=1}^k \P(x_i \notin \cs) \leq \delta$. For the second term, notice that $\abs{\cs} \geq k$ on the event $\ca^c$. Therefore, the only possibility for the algorithm to return $0$ is that the function call of \textsc{MaxHeapThreshold} on Line 10 incorrectly returns $0$. By Lemma~\ref{lemma:existing_threshold}, we have that $\P(\ce_1 \cond \ca^c)\le \delta$. This leads to the upper bound $\P(\ce_1)\le 2\delta$.
\footnotetext{The sorting algorithm of~\cite{feige1994computing} considers the setting of noisy pairwise comparisons. Since we consider noisy readings of the bits here, then each comparison in the sorting algorithm of~\cite{feige1994computing} should be done by a call to \texttt{NoisyCompareUsingReadings} (Algorithm~\ref{alg:noisycomparereading}).}
    Next, suppose that $\sfth_k({\bf x}) = 0$. Let $\ce_0=\{\widehat{\sfth}_k=1\}$ denote the error event in this case, and define the auxiliary event $\cb=\{\abs{\cs}\ge k+ \max(n\delta+n\sqrt{\delta},\frac{n}{\log n})\}$. As before, we have
    \begin{equation*}
        \P(\mathcal{E}_0)\le \P(\cb)+\P(\ce_0|\cb^c).
    \end{equation*}
    By Lemma~\ref{lemma:existing_threshold}, we know that $\P(\ce_0|\cb^c)\le \delta$. Moreover, we claim that $\P(\cb)\le 2\delta$.
    It follows that $\P(\ce_0)\le 3\delta$, and completes the proof.

    It remains to show the claim. Notice that, by Lemma~\ref{lemma:check-bit},
    \[
        \abs{\cs} \, \sim \, \mathrm{Binom}(w,1-\gamma) + \mathrm{Binom}(n-w,\gamma),
    \]
    where $w \leq k-1$ and $\gamma \leq \delta/k$. Hence, $|\mathcal{S}|$ is stochastically dominated by the random variable $k+\mathrm{Binom}(n,\delta)$. We separately consider two different cases to show the claim.

    \vspace{0.25em}
    \noindent\emph{\textbf{Case 1: $\delta\ge \frac{2}{n}$.}} In this case, we have
    \begin{align*}
        \P\left(|\mathcal{S}|\ge k+n\delta+n\sqrt{\delta}\right)&\le \P\left(\mathrm{Binom}(n,\delta)\ge n\delta+n\sqrt{\delta}\right)\\
        &\stackrel{(a)}{\le}\exp\left(-\frac{n}{2+1/\sqrt{\delta}}\right),
    \end{align*}
    where $(a)$ follows by the Chernoff bound (see, for example, Theorem 4.4 in~\cite{mitzenmacher2017probability}). To show that $\P(\cb)\le 2\delta$, it suffices to show that $\frac{n}{2+1/\sqrt{\delta}}\ge\log\frac1\delta$. This follows by the fact that $(2+1/\sqrt{\delta})\log\frac1\delta$ is maximized at $\delta=2/n$ with value $(2+1/\sqrt{2/n})\ln(n/2)$, and that $n>(2+1/\sqrt{2/n})\ln(n/2)$ for any $n\ge 1$.

    \vspace{0.25em}
    \noindent \emph{\textbf{Case 2: $\delta<\frac{2}{n}$.}} Here we assume without loss of generality that $n\ge 3$, because $\frac{n}{\log n}>n$ when $n\le 2$, and it follows that $\P(\cb)=0$. We have
    \begin{align*}
        \P\left(|\mathcal{S}|\ge k+\frac{n}{\log n}\right)&\le \P\left(\mathrm{Binom}(n,\delta)\ge \frac{n}{\log n}\right)\\
        &\stackrel{(b)}{=}\exp\left(-nD_\mathsf{KL}\left(\frac{1}{\log n}\big|\big|\delta\right)\right)\\
        &=\exp\left(-n\left(\frac{1}{\log n}\log\frac{1}{\delta\log n}+\left(1-\frac{1}{\log n}\right)\log\frac{1-1/\log n}{1-\delta}\right)\right),
    \end{align*}
    where $(b)$ follows by the Chernoff--Hoeffding theorem (see, for example, Theorem 1 in~\cite{Hoeffding1963}).
    To prove $\P(\cb)\le 2\delta$, it suffices to show that 
    \[
    n\left(\frac{1}{\log n}\log\frac{1}{\delta\log n}+\left(1-\frac{1}{\log n}\right)\log\frac{1-1/\log n}{1-\delta}\right)-\log\frac{1}{2\delta}\ge 0.
    \]
    Firstly, notice that $n\left(\frac{1}{\log n}\log\frac{1}{\delta\log n}+\left(1-\frac{1}{\log n}\right)\log\frac{1-1/\log n}{1-\delta}\right)-\log\frac{1}{2\delta}$ monotonically decreases as $\delta$ increases from $0$ to $2/n$. The proof is then completed by the fact that
    \[
    n\left(\frac{1}{\log n}\log\frac{n}{2\log n}+\left(1-\frac{1}{\log n}\right)\log\frac{1-1/\log n}{1-2/n}\right)-\log\frac{n}{4}\ge 0
    \]
    for any $n\ge 3$.
\end{proof}

\vspace{0.5em}
\emph{\textbf{Query Analysis:}}
Let $M$ denote the number of queries made by the \textsc{NoisyThreshold} algorithm. The following lemma gives an upper bound on the expectation of $M$, and hence completes the proof of Proposition~\ref{prop:upper_regime1}.
\begin{lemma} \label{lemma:number-queries-threshold}
    For any input instance $\bx$, the expected number of queries made by the \textsc{NoisyThreshold} algorithm satisfies that
    \[
    \E[M|\bx]\le\frac{n\log\frac{k}{\delta}}{\DKL}+\frac{n}{1-2p}+C\left(k+\max\left(n\delta+n\sqrt{\delta},\frac{n}{\log n}\right)\right)\log\frac{k}{\delta},
    \]
    where $C$ is a constant depending only on $p$.
    
\end{lemma}
\begin{proof}
    To bound $\E[M|\bx]$, we only need to bound the expected number of queries made by the $n$ calls of the \textsc{CheckBit} function on Line 3, and the single call of the \textsc{MaxHeapThreshold} function on Line 10.
    By Lemma~\ref{lemma:check-bit}, we know that the expected number of queries spent at Line 3 is at most
    \begin{align*}
        \frac{n}{1-2p}\left\lceil\frac{\log\frac{1-\frac{\delta}{k}}{\frac{\delta}{k}}}{\log\frac{1-p}{p}}\right\rceil &\le \frac{n}{1-2p}\left(\frac{\log\frac{k}{\delta}}{\log\frac{1-p}{p}}+1\right) \\
        &= \frac{n\log\frac{k}{\delta}}{\DKL}+\frac{n}{1-2p}.
    \end{align*}
    By Lemma~\ref{lemma:existing_threshold}, we know that the expected number of queries spent at Line 12 is at most
    \[
        C\E\left[\abs{\cs}\right]\log\frac{k}{\delta} \le C\left(k+\max\left(n\delta+n\sqrt{\delta},\frac{n}{\log n}\right)\right)\log\frac{k}{\delta},
    \]
    where the inequality holds because $\abs{\cs}\le k+\max\left(n\delta+n\sqrt{\delta},\frac{n}{\log n}\right)$ on the event that Line 12 is executed. The lemma follows by the linearity of expectation.
\end{proof}

\subsection{Regime 2: \texorpdfstring{$k> \frac{n}{\log n}$}{Lg}}
To prove Theorem~\ref{thm:threshold_achievability} in this regime,
we establish the following proposition.
\begin{proposition}
\label{prop:upper_regime2}
    Suppose $k\ge \frac{n}{\log n}$ and $k\le n/2$. There exists a variable-length algorithm, namely Algorithm~\ref{alg:checkeachbit}, that computes the $\sfth_k$ function with worst-case error probability at most $\delta$. Moreover, the number of queries $M$ made by the algorithm satisfies 
    $$\E[M|\bx]\le\frac{n\log\frac{n}{\delta}}{\DKL}+\frac{n}{1-2p}.$$
    for any input instance $\bx$.
\end{proposition}
To see that Proposition~\ref{prop:upper_regime2} implies Theorem~\ref{thm:threshold_achievability} in this regime, notice that under assumptions $k\ge \frac{n}{\log n}$ and $\delta=o(1)$, the expected query complexity of Algorithm~\ref{alg:checkeachbit} satisfies
\[
\E[M|\bx]\le\frac{n\log\frac{n}{\delta}}{\DKL}+\frac{n}{1-2p}=(1+o(1))\frac{n\log\frac{k}{\delta}}{\DKL}.
\]
Now it suffices to prove Proposition~\ref{prop:upper_regime2}.

To prove the proposition, we first analyze the error probability of Algorithm~\ref{alg:checkeachbit}. By Lemma~\ref{lemma:check-bit}, we know that $\P(\hat{x}_i\neq x_i)\le \frac{\delta}{n}$ for each $i$, where $\hat{x}_i$ is the estimate for $x_i$ in Algorithm~\ref{alg:checkeachbit}. By the union bound, we have $\P(\exists i\in[n]: \hat{x}_i\neq x_i)\le \delta$, and it follows that $\P(\mathbbm{1}_{\{\sum_{i=1}^n\hat{x}_i\ge k\}}\neq \sfth_k(\bx))\le \delta$. Therefore, Algorithm~\ref{alg:checkeachbit} has error probability at most $\delta$. 

Regarding the expected number of queries of Algorithm~\ref{alg:checkeachbit}, notice that the algorithm calls the \textsc{CheckBit} for $n$ times. By Lemma~\ref{lemma:check-bit}, the total expected number of queries is at most 
\[
 \frac{n}{1-2p}\left\lceil\frac{\log\frac{1-\frac{\delta}{n}}{\frac{\delta}{n}}}{\log\frac{1-p}{p}}\right\rceil\le \frac{n}{1-2p}\frac{\log\frac{n}{\delta}}{\log\frac{1-p}{p}}+\frac{n}{1-2p}= \frac{n\log\frac{n}{\delta}}{D_{\mathsf{KL}}(p \| 1-p)}+\frac{n}{1-2p},
\]
which completes the proof.

\subsection{Fixed-length Algorithms}\label{appd:fixed-length}
In this section, we describe a fixed-length version for each of the two proposed algorithms, and bound their number of queries and worst-case error probability.

\subsubsection{Fixed-length Version of Algorithm~\ref{alg:proposed}}
Notice that in Algorithm~\ref{alg:proposed}, all the queries are spent on the $n$ executions of the \textsc{CheckBit} function on Line 3 and the call of the \textsc{MaxHeapThreshold} function on Line 12. Notice that in Lemma~\ref{lemma:existing_threshold}, a deterministic upper bound is established for the number of queries of the \textsc{MaxHeapThreshold} function. Thus, the randomness in the proposed algorithm only comes from the number queries of the \textsc{CheckBit} function. The main idea in this the construction of the fixed-length algorithm is to convert the \textsc{CheckBit} algorithm into one with bounded second moment for the number of queries. More specifically, we define the \textsc{SafeCheckBit} algorithm in the following two step procedure:
\begin{enumerate}
    \item Run the \textsc{CheckBit} algorithm until it halts, or it is about to make the $(\eta\log \eta)$-th query, where $\eta\triangleq\frac{1}{1-2p}\left\lceil\frac{\log\frac{1-\delta}{\delta}}{\log\frac{1-p}{p}}\right\rceil$.
    \item If the \textsc{CheckBit} algorithm halts, then return the algorithm output, otherwise, restart the \textsc{CheckBit} algorithm.
\end{enumerate}
The construction of the safe algorithm is first proposed in \cite{gu2023optimal}. It is shown in Corollary 19 of~\cite{gu2023optimal} that the error probability of the \textsc{SafeCheckBit} algorithm is at most $(1+o(1))\delta$. Moreover, its number of queries $M_s$ satisfies
\begin{equation}
\label{eq:safe-bound}
\E[M_s]\le\frac{\eta}{1-\frac{1}{\log \eta}}\;\text{ and }\;\mathsf{Var}(M_s)=O\left(\log^3\frac{1}{\delta}\right).
\end{equation}
In the proposed fixed-length algorithm, we replace the call of \textsc{CheckBit}$(x_i,\frac{\delta}{k},p)$ on Line 3 of Algorithm~\ref{alg:proposed} by \textsc{SafeCheckBit}$(x_i,\frac{\delta}{k},p)$. 
Moreover, we add a termination rule in the for loop from Line 2 to Line 6 that the algorithm terminates and declares failure if it is about to use the $$\left(\frac{n\eta'}{1-\frac{1}{\log \eta'}}+n\sqrt{\log\frac{k}{\delta}}\right)\text{-th}$$ query, where $\eta'\triangleq \frac{1}{1-2p}\left\lceil\frac{\log\frac{k-\delta}{\delta}}{\log\frac{1-p}{p}}\right\rceil$.

With the termination rule, it readily follows that the number of queries made by the algorithm is always at most $(1+o(1))\frac{n\log\frac{k}{\delta}}{\DKL}$. 
Let $X$ denote the total number of queries made by the $n$ executions of the \textsc{SafeCheckBit} algorithm. 
Because the number of queries made by each execution of the \textsc{SafeCheckBit} algorithm is independent of one another, we have $$\E[X]\le\frac{n\eta'}{1-\frac{1}{\log \eta'}}\;\text{ and }\;\mathsf{Var}(X)=O\left(n\log^3\frac{k}{\delta}\right).$$ By Chebyshev's inequality, we have
\begin{equation*}
    \P\left(X\ge \frac{n\eta'}{1-\frac{1}{\log \eta'}}+n\sqrt{\log\frac{k}{\delta}}\right)\le O\left(\frac{n\log^3\frac{k}{\delta}}{n^2\log\frac{k}{\delta}}\right)=O(n^{-1+o(1)})=o(\delta),
\end{equation*}
where the last two equalities hold true under the assumption that $\delta\ge n^{-1+\eps}$ for some positive constant $\eps$. Hence, by the union bound and Lemma~\ref{lemma:error-prob-threshold}, the error probability of the proposed fixed-length algorithm is at most $(3+o(1))\delta$. This proves Corollary~\ref{cor:fixed-length} for the regime $k\le\frac{n}{\log n}$.

\subsubsection{Fixed-length Version of Algorithm~\ref{alg:checkeachbit}}
To construct the fixed-length version of Algorithm~\ref{alg:checkeachbit}, we similarly replace the \textsc{CheckBit} function by the \textsc{SafeCheckBit} function and add a termination rule based on the number of queries spent. More specifically, we replace the call of function \textsc{CheckBit}$(x_i,\delta/n)$ on Line 3 Algorithm~\ref{alg:checkeachbit} by \textsc{SafeCheckBit}$(x_i,\delta/n)$. Furthermore, we add the termination rule in the for loop that the algorithm terminates and declares failure if it is about to use the
$$\left(\frac{n\eta''}{1-\frac{1}{\log \eta''}}+n\sqrt{\log\frac{n}{\delta}}\right)\text{-th}$$ query, where $\eta''\triangleq \frac{1}{1-2p}\left\lceil\frac{\log\frac{n-\delta}{\delta}}{\log\frac{1-p}{p}}\right\rceil$.

By the termination condition, we know that the algorithm deterministically uses at most $(1+o(1))\frac{n\log\frac{n}{\delta}}{\DKL} =(1+o(1))\frac{n\log\frac{k}{\delta}}{\DKL}$ queries, where the equality follows because $k>\frac{n}{\log n}$. By an analogous analysis as in the previous section, we know that the error probability of the algorithm is at most $(1+o(1))\delta$ given that $\delta\ge n^{-1+\epsilon}$. This completes the proof for the regime $k>\frac{n}{\log n}$.

\section{Implementation Details} \label{sec:implementation}
In this part, we give some details about the simulation results shown in Figure~\ref{fig:comparison}. In this simulation, our algorithm for computing the $\mathsf{MAJORITY}$ function (i.e., the threshold function with $k=n/2$) is compared to the existing algorithm proposed in~\citet{feige1994computing}. A binary sequence of length $n=100$ chosen uniformly at random is inputted to the algorithms, and the desired error probability in the algorithms is set to $\delta = 10^{-2}$.
For a sequence generated uniformly at random, the law of large numbers implies that the fraction of ones in the sequence is approximately $1/2$ with high probability. This scenario is the most likely to induce errors for an algorithm computing the $\mathsf{MAJORITY}$ function, making it critical for assessing the worst-case performance of the algorithm.
For $k=n/2$, the existing work by~\citet{feige1994computing} proposes a \emph{noisy sorting} algorithm that sorts the input binary sequence, and returns the $k$-th largest element in the sorted sequence as an estimate of the threshold function. The reader is referred to Section 3 in~\citet{feige1994computing} for more details about the existing algorithm. For the simulation of our proposed algorithm, we use Algorithm~\ref{alg:checkeachbit} (since $k=n/2$). At each value of $p$ ranging from 0.01 to 0.25, we run 10000 realizations of input sequences and compute the average number of queries used by the two algorithms. 

\end{document}